\documentclass[journal]{IEEEtran}
\usepackage{color}
\usepackage{glossaries}
\usepackage[printonlyused]{acronym} 
\usepackage{amsmath}
\usepackage[caption=false,font=footnotesize]{subfig}
\usepackage{bm,amssymb}
\usepackage{amsthm}
\usepackage[noadjust]{cite}
\usepackage{pifont}
\theoremstyle{plain}
\newtheorem{remark}{Remark}
\DeclareMathAlphabet\mathbfcal{OMS}{cmsy}{b}{n} 
\usepackage{tcolorbox}
\usepackage{enumitem}
\usepackage{graphicx}
\usepackage[normalem]{ulem}
\usepackage{multirow}

\usepackage{tikz}
\usepackage[utf8]{inputenc}
\usepackage{pgfplots} 
\usepackage{pgfgantt}
\usepackage{pdflscape}
\pgfplotsset{compat=newest} 
\pgfplotsset{plot coordinates/math parser=false}

\newcommand\blue[1]{{\color{black}#1}}

\newcommand\fuxi[1]{{\color{black} #1}}

\acrodef{GPS}{Global Positioning System}
\acrodef{LOS}{line-of-sight}
\acrodef{NLOS}{non-line-of-sight}
\begin{document}
 
\title{
5G Positioning and Mapping with Diffuse Multipath
}

\author{Fuxi Wen,~\IEEEmembership{Senior Member,~IEEE}, Josef Kulmer, Klaus Witrisal and Henk Wymeersch,~\IEEEmembership{Senior Member,~IEEE} \thanks{F.~Wen and H.~Wymeersch are with the Department of Electrical Engineering, Chalmers University of Technology, Sweden. J.~Kulmer and K.~Witrisal are with the Signal Processing and Speech Communication Laboratory (SPSC) of Graz University of Technology, Austria. This work was supported, in part, by the Marie Skłodowska-Curie grant agreement No 700044 and by the Swedish Research Council under grant 2018-03701.}
}
\markboth{\today}%
{Shell \MakeLowercase{\textit{et al.}}: Bare Demo of IEEEtran.cls for IEEE Journals}
\maketitle
 
\begin{abstract}
5G mmWave communication is useful for positioning due to the geometric connection between the propagation channel and the propagation environment. Channel estimation methods can exploit the resulting sparsity to estimate parameters (delay and angles) of each propagation path, which in turn can be exploited for positioning and mapping. When paths exhibit significant spread in either angle or delay, these methods break down or lead to significant biases. We present a novel tensor-based method for channel estimation that allows estimation of mmWave channel parameters in a non-parametric form. The method is able to accurately estimate the channel, even in the absence of a specular component. This in turn enables positioning and mapping using only diffuse multipath. Simulation results are provided to demonstrate the efficacy of the proposed approach. 
\end{abstract}


\section{Introduction}
5G mmWave signals present unique opportunities for positioning or user devices, due to their large bandwidths, arrays with many antenna elements and favorable propagation conditions \cite{wymeersch20175g}. 5G mmWave is currently a study item for 3GPP-R17 and has the potential not only to provide performance better than \ac{GPS}, but also enable precise orientation estimation. Moreover, due to the high degree of resolvability of propagation paths, multipath information can naturally be exploited, both for positioning as well as for mapping of the environment \cite{WitrisalSPM16} 
Applications of 5G mmWave positioning include traditional emergency call localization and personal navigation, but also more disruptive topics such as localization or robots and autonomous vehicles, as well as augmented and virtual reality applications. 

In order to develop a localization method, an understanding of the mmWave channel is needed. mmWave propagation, occurring at carrier frequencies above 24 GHz, has been shown to be characterized by limited scattering, no diffraction and shadowing, and the existence of only a few propagation paths. Each of the paths is thus largely determined by the propagation environment and characterized by channel gains, angles of arrival, angles of departure, and delays.  Propagation paths may be of a deterministic specular nature, when the surface on which waveforms impinge is sufficiently smooth, or of a stochastic diffuse/scattering nature when the surface is relatively rough, or a combination of both. Hence, in general, each path (except the \ac{LOS} path) is in fact a cluster of paths, with similar angles and delays \cite{akdeniz2014millimeter}. When the paths within a cluster are not resolvable in either angles or delays, they lead to fluctuations in the received power. This is the model typically assumed in the communication literature.  On the other hand, when intra-cluster paths are resolvable, they should be properly estimated in order to avoid biasing the estimation of angles and delays. 

A cluster can be characterized in multiple ways. Traditionally, a statistical model has been considered, whereby a cluster is modeled though a mean and a spread in both angle and delay domain \cite{fleury2000first}. Given such a model, there is a rich literature on second-order estimation methods that are able to accurately and blindly estimate the mean and spread of a cluster \cite{besson2000decoupled,yucek2008time}.
The models for spatially distributed sources have been classified into two types, namely incoherently distributed (ID) sources and coherently distributed (CD) sources. On one hand, for ID sources, signals coming from different points of the same distributed source can be considered as uncorrelated \cite{Shahbazpanahi2001,Li2007,Zoubir2008,Dai2017}. On the other hand, in the scenario of CD sources, the received signal components are delayed and scaled replicas from different points within the same source \cite{Lee2003,Zoubir2008b,Zhou2017}. 
In \cite{Yan2018}, the performance bound  is studied of the tracking accuracy
in sparse mmWave channels that includes cluster angular
 spreads 
However, while such subspace methods are powerful, in the context of mmWave communication, the signal structure and presence of dedicated pilot signals should be exploited to develop faster methods. There is thus a lack of first-order methods for quickly estimating channel parameters and their spread. This explains why 5G mmWave localization has considered either only the \ac{LOS} path, or treated multipath as purely specular \cite{shahmansoori2018position}. 
Standard 5G mmWave channel estimation is based on either compressive sensing approaches \cite{alkhateeb2014channel}, which express the sparsity in an appropriate domain, \fuxi{or on tensor decompositions, where the dominant higher-order singular values can be related to the dominant signal paths \cite{zhou2016channel,Zhou2017tensor}. 
A joint tensor decomposition and compressed sensing based multidimensional channel parameter estimation method is proposed in \cite{Ruble2020}. However,}
these methods do not account for the intra-cluster spread of angles or delay. 

In this paper, we propose a tensor-based method for estimating a 5G mmWave channel in terms of the angles and delays of the individual paths within each \ac{NLOS} cluster. The method makes no a priori assumption regarding the number of paths per cluster. \fuxi{The problem of clustering is not our focus, and standard clustering methods can be applied, such as $k$-means and 
density-based spatial clustering of applications with noise (DBSCAN) \cite{Martin1996}}. Following a clustering of paths, the statistics of each cluster can be determined, which are finally fed to a positioning and mapping method. 
The proposed method is able to determine the dominant clusters and accurately estimate the cluster statistics, even for clusters that have no specular component. Building on this, we present a positioning and mapping method that accurately localizes the user and maps the environment by exploiting the diffuse multipath, rather than considering it as a disturbance. Our main contributions are the following:
\begin{itemize}
    \item We derive a novel method  for estimating mmWave channels in the presence of combined specular and scattered components, based on a tensor decomposition.
    \item We provide a detailed evaluation of the proposed method in a three-dimensional propagation environment, demonstrating its performance under varying levels of surface roughness. 
    \item We propose a 5G mmWave localization and mapping method that is able to operate in the absence of \ac{LOS} and specular multipath. The method utilizes only the diffuse multipath for positioning and mapping. 
\end{itemize}



\section{Tensors and Tensor Operations}

\subsection{Definitions and Notations}
The tensor operations used in this paper are consistent with \cite{Haardt2008}.
An $R$-D tensor is denoted by $\mathbfcal{A}\in \mathbb{C}^{M_1 \times M_2 \times \cdots \times M_R}$, where $M_r$ is the size of the $r$th mode of the tensor and $R \ge 3$.
We use $a_{m_1, m_2, \cdots, m_R}$ to  represent the $(m_1, m_2, \cdots, m_R)$ entry.\\
\emph{Unfolding:} The $r$-mode unfolding of $\mathbfcal{A}$ is written as $\mathbf{A}_{(r)} \in \mathbb{C}^{M_r \times (M_1\cdots M_{r-1}M_{r+1} \cdots M_R)}$ where the order of the columns is chosen according to \fuxi{Definition 1 in \cite{Lieven2000}}.\\
\emph{Product:} The $r$-mode product of a tensor $\mathbfcal{A} \in \mathbb{C}^{M_1 \times M_2 \times \cdots \times M_R}$ and a matrix $\mathbf{U} \in \mathbb{C}^{N_r \times M_r}$ along the $r$th mode is denoted as \fuxi{Definition 8 in \cite{Lieven2000},} 
\begin{equation} 
\label{eq1} 
\mathbfcal{B}  =  \mathbfcal{A} \times_r \mathbf{U} \in \mathbb{C}^{M_1 \times \cdots \times M_{r-1} \times N_r \times M_{r+1} \times \cdots \times M_R}.
\end{equation}
\emph{Concatenation:} \fuxi{We use the operator $\left[\mathbfcal{A}_1 \sqcup_{R+1} \mathbfcal{A}_2\right] \in \mathbb{C}^{M_1 \times M_2 \times \cdots \times M_{R} \times 2}$ to represent the concatenation of two tensors  $\mathbfcal{A}_1 \in \mathbb{C}^{M_1 \times M_2 \times \cdots \times M_{R}}$ and $\mathbfcal{A}_2 \in \mathbb{C}^{M_1 \times M_2 \times \cdots \times M_{R}}$, along the $(R$+$1)$th mode \cite{Roemer2007}}.

\subsection{Tensor Decompositions} 
 There exist various decompositions of tensors and definitions of the rank of a tensor. We consider here the  CANDECOMP/PARAFAC (CP) decomposition and the Tucker decomposition. \\
\emph{CP decomposition} decomposes an $R$-D tensor $\mathbfcal{X}$ as a sum of rank-one tensors
\begin{align}
\mathbfcal{X}    = \sum_{d=1}^D \gamma_d \mathbf{a}^{(1)}_d \circ \mathbf{a}^{(2)}_d \ldots \circ \mathbf{a}^{(R)}_d,
\end{align}
where $\circ$ denotes outer product.
The rank $D$ of a tensor is defined as the smallest number of rank one tensors that generate $\mathbfcal{X}$ as their sum.  
In other words, it is the smallest number of components in an exact CP decomposition \cite{daCosta2011,Liu2016}. 
 The $r$-rank of a tensor is the column rank of $\mathbfcal{X}_{(r)}$ \cite{Yokota2017}. 
\\
\emph{Tucker decomposition} is a form of higher-order principal component analysis.
It decomposes a tensor into a core tensor 
multiplied (or transformed) by a matrix along each mode.  The matrix can be thought of as the principal components in each mode.
\section{System Model}

\begin{figure}
 \centering
 \includegraphics[width=0.9\columnwidth]{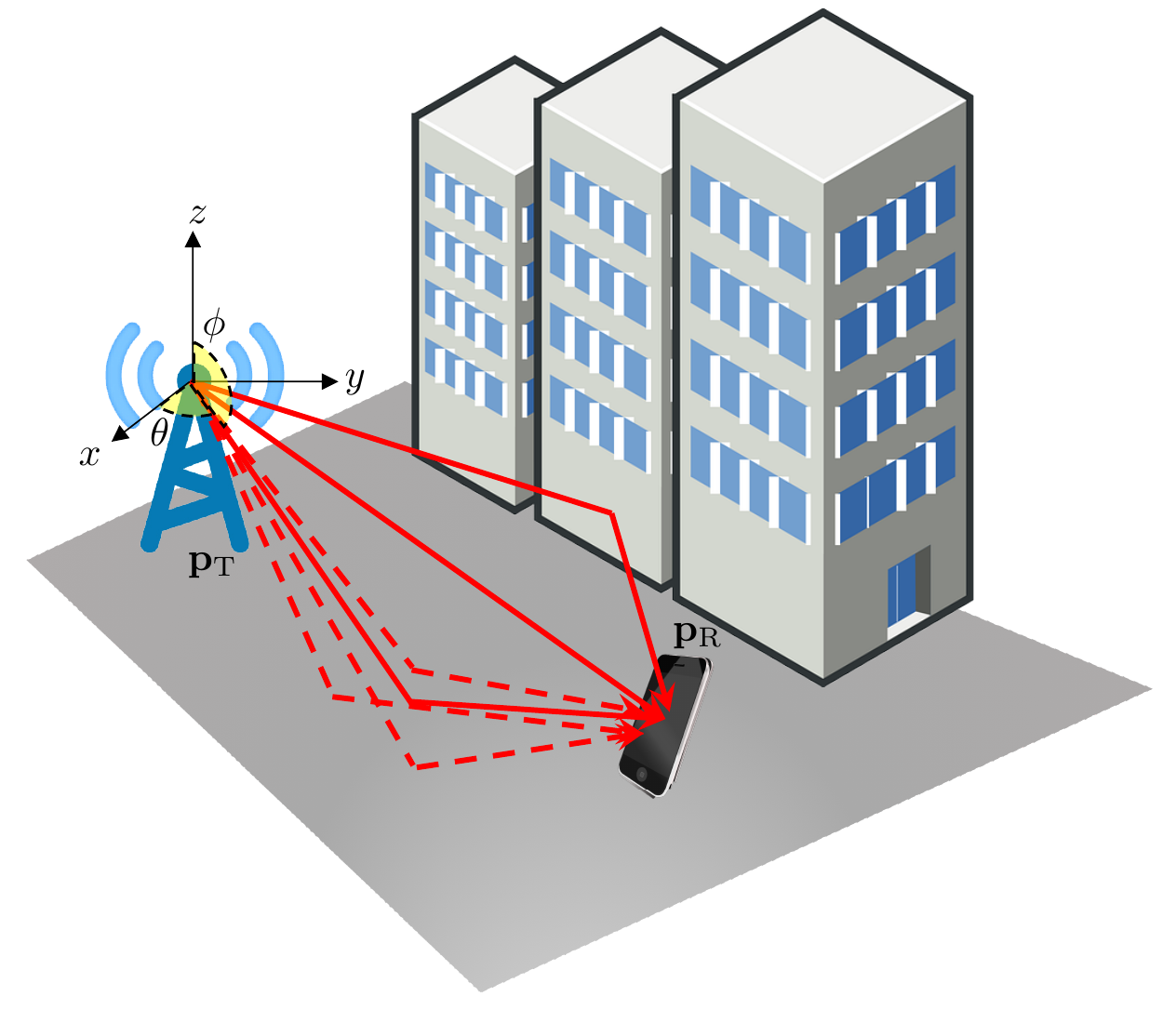}
 \caption{Illustration of the considered scenario with 1 LOS path and 2 NLOS clusters (i.e.,  $K=3$).}
 \label{fig:scenario}
 \end{figure}

We consider a $3$-dimensional (3$D$) scenario with a single 5G transmitter with known location $\bm{p}_{\text{T}}$ and orientation, a receiver with unknown location  $\bm{p}_{\text{R}}$, and a physical propagation environment, characterized by surfaces, as depicted in Figure~\ref{fig:scenario}. 
The transmitter and receiver both employ uniform rectangular arrays (URAs) consist of sensors in a grid of size $M_{\rm{T}}= M_1 \times M_2$ and $M_{\rm{R}}= M_3 \times M_4$, and exchange MIMO-OFDM signals with $M_5$ sub-carriers 
and sub-carrier spacing  $\Delta_f$. 
The received signal on subcarrier $i$ is of the form
\begin{align}
    \mathbf{Y}_i=\mathbf{H}_i \mathbf{S}_i + \mathbf{N}_i, 
\end{align}
where $\mathbf{S}_i$ is a known pilot signal with orthogonality property ($\mathbf{S}_i \mathbf{S}_i^{\mathsf{H}}$ is a scaled identity matrix) and $\mathbf{N}_i$ is i.i.d.~Gaussian noise. 
Then we have
\begin{align}
    \mathbf{Y}_i \mathbf{S}_i^{\mathsf{H}} =\mathbf{H}_i + \mathbf{N}_i\mathbf{S}_i^{\mathsf{H}}.
\end{align}

\fuxi{For subcarrier $i$, we receive $\mathbf{Y}_i$, which is an {$M_3M_4 \times M_1M_2$} matrix. Then we convert these $M_5$ matrices (one per subcarrier) in a 5D tensor of suitable dimension, $\mathbfcal{Y} \in \mathbb{C}^{M_1 \times M_2 \times M_3 \times M_4 \times M_5}$.}
The channel matrix $\mathbf{H}_i$ depends on the array structure and the propagation environment, described next.
Our aim is to determine $\bm{p}_{\text{R}}$ and map the propagation environment. 

\subsection{Array Steering Vector for URA}
\label{sec:steering_vectors}
The transmit and receive arrays are planar arrays, comprising omni-directional elements on a uniform grid of rectangular shape with inter-element spacing  equal to half of the signal's wavelength. 
Transmit and receive URAs consist of sensors are indexed by $(m_1,m_2)$ and $(m_3,m_4)$, respectively. 

The URA steering vector corresponding to the $l$th source can be formed as
\begin{equation}
    \mathbf{a}\left(\omega_{l,1},\omega_{l,2}\right) = \mathbf{a}\left(\omega_{l,1}\right) \otimes \mathbf{a}\left(\omega_{l,2}\right),
\end{equation}
where $\otimes$ is Kronecker product, 
$\mathbf{a}\left(\omega_{l,1}\right) =  
      [ a_{1}(\omega_{l,1}) \, \cdots \, a_{m_1}(\omega_{l,1}) \, \cdots \, a_{M_1}(\omega_{l,1}) ]^T$ and $\mathbf{a}(\omega_{l,2}) = [a_{1}(\omega_{l,2})\, \cdots \, a_{m_2}(\omega_{l,2})\, \cdots \, a_{M_2}(\omega_{l,2})]^T$ are equivalent to the uniform linear array steering vectors 
composed of $M_1$ and $M_2$ sensors lying on $y$-axis and $z$-axis, respectively. The first sensor is taken as the reference sensor so that (up to a global phase) 
\begin{equation}
\label{eq_amomega}
a_m(\omega) = e^{j (m-1) \omega}.
\end{equation}
The spatial frequencies associated with the azimuth $\theta_l$ and
elevation angle $\phi_l$ of the $l$th source follow as
\begin{equation} 
        \omega_{l,1}  = \pi \sin(\theta_l)\sin(\phi_l), 
        \quad
        \omega_{l,2}  =  \pi \cos(\phi_l).
\end{equation}
\subsection{Channel Model}
%
%
%
We propose a generative model for simulating the diffuse multipath of mmWave channels, based on \cite{kulmer2018impact,degli2007measurement}. This model starts from generating points on the surface, based on the its roughness. Then, for each point, the channel parameters are computed (angles, delay, gains). Finally, the model is expressed in a tensor representation. For smooth reflective surfaces, the model reverts to the one used in \cite{shahmansoori2018position}. 

\subsubsection{Surface Roughness and Scattering}
The propagation environment consists of $K$ well-separated clusters, each cluster $k$ corresponds to a physical object (e.g., a wall, a ground reflection), \fuxi{described by MPCs,}
characterized by two parameters \cite{kulmer2018impact,degli2007measurement}: 
\begin{itemize}
 \item The scattering coefficient $S\in [0,1]$, which quantifies the relative amount ({with respect to absorption}) of total scattered amplitude, and was identified to be $S \ge 0.4$ \cite{degli2007measurement,jarvelainen2014sixty}. 
 \item The directivity parameter $\alpha_R \ge 0$ which describes the width of the scattering lobe originating at the reflective surface. At rough surfaces (in comparison to the signal's wavelength), the scattering power has a large intra-cluster spread, corresponding to a small directivity $\alpha_R \rightarrow 0$. At smooth surfaces, the spread of scattering power is reduced, equivalent to more directivity $\alpha_R \to \infty$. Hence, $\alpha_R$ may be associated to surface roughness. Typical values are in a range of $\alpha_R\in\{1,\ldots,11\}$ \cite{degli2007measurement,jarvelainen2014sixty}.
\end{itemize}
Combined, $\alpha_R$ and $S$ can be used to determine the cluster power and cluster spread through the joint angular delay power spectrum (JADPS) which describes the scattered power $p_{\text{DM}}(\bm{p})$ from any point $\bm{p}$ \cite{kulmer2018impact}. 
Cluster $k$ gives rise to $L_k$ scatter points,
where the total number of paths is
{$P = \sum_{k=1}^{K}L_k$}. 
For the LOS path, $L_k=1$. Each scatter point $\bm{p}_{kl} \in \mathbb{R}^3$ 
lies on the $k$-th surface with scatter point index $0<l\leq L_k$. 

\subsubsection{Generation of Channel Parameters}
Given a path between $\bm{p}_{\rm{R}}$ and $\bm{p}_{\rm{T}}$ via $\bm{p}_{kl}$, the path delay $\tau_{kl}$, as well as azimuth and elevation angles of the angle-of-departure (AOD)
$(\theta_{kl},\phi_{kl})$  and  of the angle-of-arrival (AOA)
$(\vartheta_{kl},\varphi_{kl})$ follow from standard geometry and can be found in the Appendix~\ref{sec:appGeometry}. 
%
%
%
Finally, each path from a scatter point has a gain $\gamma_{kl}$, which we propose to comprise a constant amplitude per cluster and a random phase, uniform over $[0,2\pi)$. Motivation and additional details of this model are provided in Appendix \ref{sec:gensp}.

%
%

\subsubsection{Tensor Formulation}\label{Sec:TF}
Let
\begin{equation} 
        \omega_{kl,1}  = \pi \sin(\theta_{kl})\sin(\phi_{kl}), 
        \quad
        \omega_{kl,2}  =  \pi \cos(\phi_{kl}),
\end{equation}
and
\begin{equation} 
        \omega_{kl,3}  = \pi \sin(\vartheta_{kl})\sin(\varphi_{kl}), 
        \quad
        \omega_{kl,4}  =  \pi \cos(\varphi_{kl}),
\end{equation} 
the channel response in frequency domain for sub-carrier $i$ with frequency $f_i$ is represented as \cite{Sha2019}
\begin{align}
\mathbf{H}_i = \sum_{k = 1}^{K} \sum_{l = 1}^{L_k} \gamma_{k l} e^{-j 2\pi f_i \tau_{k l}}  \mathbf{a}_{\rm{R}} \left( \vartheta_{kl},\varphi_{kl} \right) \mathbf{a}^{\mathsf{H}}_{\rm{T}} \left( \theta_{kl},\phi_{kl} \right),
\end{align}
where
\begin{equation} 
\mathbf{a}_{\rm{T}}\left(\omega_{kl,1},\omega_{kl,2}\right) = \mathbf{a}\left(\omega_{kl,1}\right) \otimes \mathbf{a}\left(\omega_{kl,2}\right) \in \mathbb{C}^{M_T \times 1},
\end{equation} 
and
\begin{equation} \mathbf{a}_{\rm{R}}\left(\omega_{kl,3},\omega_{kl,4}\right) = \mathbf{a}\left(\omega_{kl,3}\right) \otimes \mathbf{a}\left(\omega_{kl,4}\right) \in \mathbb{C}^{M_R \times 1}.
\end{equation} 

\fuxi{For subcarrier $i$, $\mathbf{H}_i$ is an {$M_3M_4 \times M_1M_2$} matrix. Then we convert these $M_5$ matrices (one per subcarrier) in a 5D tensor of suitable dimension, $\mathbfcal{H} \in \mathbb{C}^{M_1 \times M_2 \times M_3 \times M_4 \times M_5}$.}




\section{Proposed Method}
We now present our method for localizing the receiver and the cluster locations. 

 \subsection{Tensor Representation}\label{Sec:CM}
 The $(m_1,m_2, m_3, m_4, m_5)$ entry of the channel response in frequency domain $\mathbfcal{H}\in \mathbb{C}^{M_1 \times M_2 \times M_3 \times M_4 \times M_5}$ is described as
\begin{align}
 h_{m_1 m_2 m_3 m_4 m_5} = & \sum_{k=1}^{K}  \sum_{l=1}^{L_k} \gamma_{kl} \nonumber a_{m_1}\left(\omega_{kl,1}\right) a_{m_2}\left(\omega_{kl,2}\right) \\ &    a_{m_3}\left(\omega_{kl,3}\right) a_{m_4}\left(\omega_{kl,4}\right) a_{m_5}(\omega_{kl,5}), 
\end{align}
where the spatial frequency 
$\omega_{kl,5} = 2\pi\Delta_{f} \tau_{kl}$, and $a_{m}(\omega)$ is defined in (\ref{eq_amomega}).
The response can be described as a CP model (sum of {\color{blue}$P$} rank-one tensors),  
\begin{equation}
\label{eq4H}
   \mathbfcal{H} = \sum_{p=1}^{P} \gamma_{p} \mathbf{a}_{p,1} \circ \mathbf{a}_{p,2} \circ \mathbf{a}_{p,3} \circ \mathbf{a}_{p,4} \circ
   \mathbf{a}_{p,5} .
\end{equation}
For $r = 1, 2, \cdots, 5$,
\begin{equation}
 \mathbf{a}_{p,r} = \begin{bmatrix}
       a_{1}(\omega_{p,r})  &  a_{2}(\omega_{p,r})  & \cdots   & a_{M_r}(\omega_{p,r})
     \end{bmatrix}^{\mathsf{T}}.
\end{equation}
The array manifold for the $r$th dimension is defined as
\begin{equation}
    \mathbf{A}_r = \begin{bmatrix}
       \mathbf{a}_{1,r} & \cdots & \mathbf{a}_{p,r} & \cdots & \mathbf{a}_{P,r}
     \end{bmatrix} \in \mathbb{C}^{M_r \times P}.
\end{equation}

For multiple measurement scenarios, the augmented observation tensor is described as
\begin{equation}
\mathbfcal{Y} = \big[\underbrace{\mathbfcal{H} \sqcup_{6} \cdots \mathbfcal{H}}_{M_6}\big] + \mathbfcal{N} \in \mathbb{C}^{M_1 \times M_2 \times M_3 \times M_4 \times M_5 \times M_6},
\end{equation}
where $M_6$ is the subsequent time instants, $\mathbfcal{N}$ is the noise tensor.

 \subsection{Multipath Components (MPC) Parameter Estimation}


\subsubsection{Estimate the number of paths $P$}

To estimate geometrical parameters such as AOD, AOA and delay, the first step is to estimate the number {\color{blue}$\hat{P}$} of signal components in \eqref{eq4H}.
In the CP model, a tensor is decomposed into a sum of rank-one  tensors, which are expressed as the outer product of vectors. In practice, each rank-one component corresponds to a natural source or signal. Finding the tensor rank or number of multilinear components in the underlying CP model of noisy tensor observations is an important research topic. Existing approaches to CP rank estimation from noisy observations include \cite{Liu2016}.

$R$-D minimum description length (MDL) \cite{Yokota2017} is utilized for tensor rank estimation, which is proposed by stacking the measurement tensor into a matrix with the $r$-mode unfolding operation, 
\begin{equation}
  \mathbfcal{Y} \xrightarrow[\text{unfolding}]{r\text{-mode}}  \mathbf{Y}_{(r)} .
\end{equation}

The eigenvalue spectrum $\mathbf{\Lambda}_{r}$ obtained from the singular value decomposition (SVD) of $\mathbf{Y}_{(r)}$ and MDL are used for $r$-rank $\hat{P}_r$ estimation, 
\begin{equation}
\mathbf{Y}_{(r)} \xrightarrow{\text{SVD}}  \mathbf{\Lambda}_{r}  \xrightarrow{\text{MDL}} \hat{P}_r.
\end{equation}
After obtaining $r$-rank, the tensor rank is estimated as
\begin{equation}
\hat{P} = \max{\{\hat{P}_1, \hat{P}_2, \cdots, \hat{P}_R\}},
\end{equation}
to ensure a high number of estimated paths, required for cluster mean and cluster spread estimation. 
In general, $\hat{P}_r\ll P$, so the rank is always underestimated.


\subsubsection{Angle and Delay Estimation}

After estimating the number of resolvable signal components $\hat{P}$, an $R$-D subspace is obtained via CP Decomposition \cite{Kolda2009}.
For URA, tensor or $N$-D ESPRIT \cite{Haardt2008,RoemerHaardtGaldo2014,Sahnoun2017a} is applied for channel parameter estimation. 
Let $\mathbf{U}_{r} \in \mathbb{C}^{M_r \times \hat{P}}$ be the subspace spanned by $\mathbf{A}_{r}\in \mathbb{C}^{M_r \times \hat{P}}$, which is obtained by applying CP decomposition on $\mathbfcal{Y}$. 
The main idea of tensor-ESPRIT  is exploiting the multidimensional shift invariance property of the measurements. 
For each dimension, the array is divided into two subarrays with same number of elements. 
The subarrays may overlap and an element may be shared by the two subarrays. 
For the $r$th dimension, we have
\begin{equation}
\label{eqAU}
\mathbf{A}_{r} = \mathbf{U}_{r} \mathbf{D}_r, 
\end{equation}
where $\mathbf{D}_r \in \mathbb{C}^{\hat{P} \times \hat{P}}$ is a non-singular matrix.
We further define two sub-matrices,
\begin{equation}\label{eq2-13}
\mathbf{U}_{1,r} = \mathbf{J}_{1,r}^{(n)}\mathbf{U}_{r}  \text{ and }
\mathbf{U}_{2,r} = \mathbf{J}_{2,r}^{(n)}\mathbf{U}_{r},
\end{equation}
where $\mathbf{J}_{1,r}$ and $\mathbf{J}_{2,r}$ are two selection matrices,
\begin{eqnarray}\label{eq2-14}
\mathbf{J}_{1,r}^{(n)} &=& \left[ \mathbf{I}_{M_r-n} \quad \mathbf{0}_{(M_r-n) \times n}\right], \nonumber\\
\mathbf{J}_{2,r}^{(n)} &=& \left[  \mathbf{0}_{(M_r-n) \times n} \quad \mathbf{I}_{M_r-n} \right],
\end{eqnarray}
where $\mathbf{I}_n$ denotes identity matrix of size $n \times n$ and $\mathbf{0}_{m \times n}$ denotes zero matrix of size $m \times n$.
For convenience, we focus on $n=1$, $\mathbf{J}_{1,r}^{(n)}$ and $\mathbf{J}_{2,r}^{(n)}$ are simplified as $\mathbf{J}_{1,r}$ and $\mathbf{J}_{2,r}$.
Then we have
\begin{equation}\label{eq2-15}
\mathbf{J}_{1,r}\mathbf{A}_{r}  = \mathbf{J}_{2,r}\mathbf{A}_{r} \mathbf{\Phi}_r,
\end{equation}
where
\begin{equation}
\mathbf{\Phi}_r = \textrm{diag}\begin{bmatrix}
         e^{-j \omega_{1,r}}& e^{-j \omega_{2,r}} & \cdots &  e^{-j \omega_{\hat{P},r}}
     \end{bmatrix}.
\end{equation}
Substituting (\ref{eqAU}) and (\ref{eq2-13}) into (\ref{eq2-15}), we have
\begin{equation}
\mathbf{U}_{1,r} = \mathbf{U}_{2,r} \mathbf{\Psi}_r,
\end{equation}
where
\begin{equation}
\mathbf{\Psi}_r = \mathbf{D}_r \mathbf{\Phi}_r \mathbf{D}_r^{-1} \in \mathbb{C}^{\hat{P} \times \hat{P}}.
\end{equation}

The equations in  (\ref{eq2-15}) are over-determined.  The simplest choice to estimate $\mathbf{\Psi}_r$ is using the least squares (LS) method and the resulting closed-form solution is given by
\begin{equation}\label{eq2-16}
\hat{\mathbf{\Psi}}_r = \left( \mathbf{U}_{2,r} \right)^{\dagger}\mathbf{U}_{1,r},
 \end{equation} 
where $\dagger$ denotes the Moore-Penrose matrix inverse.
Let $\lambda_{1,r},  \lambda_{2,r}, \cdots, \lambda_{\hat{P},r}$ be the eigenvalues of $\hat{\mathbf{\Psi}}_r$, the mode $r$ frequencies are estimated by using
\begin{equation}\label{eq2-17}
\omega_{p,r} =  -\angle\left( \lambda_{p,r}\right), p = 1, 2, \cdots, \hat{P},
\end{equation}
where $\angle( \cdot )$ denotes the argument of a complex number. 

\begin{remark}
 Beam-space tensor ESPRIT can be applied for hybrid URA structure \cite{Wen2018GC} and beam-space tensor MUSIC is applicable for hybrid arbitrary array geometry \cite{Zhou2017}
 \end{remark}
 
\subsubsection{Clustering the MPCs} \label{sec:clustering}

Clustering techniques, such as $k$-means are applied to group the 5-D parameters of the estimated $\hat{P}$ multi-path components $\bm{\omega}_{p} = \begin{bmatrix}
    \omega_{p,1}    &    \omega_{p,2}  &  \cdots  &  \omega_{p,5} 
     \end{bmatrix}$.
It can be extended to other techniques such as connectivity-based, distribution-based and density-based \cite{Jain2010}. Given a set of estimates $\{\bm{\omega}_{p}, p = 1, 2, \cdots, \hat{P}\}$, our objective is to partition the data set into $K$ clusters, we assume that the value of $K$ is given or can be estimated from model order selection techniques \cite{Bishop2006}. 
\fuxi{Recently, the challenges and opportunities in clustering-enabled wireless channel modeling were discussed in \cite{He2018}. A framework of automatic clustering and tracking algorithm was proposed for the MPCs in time-variant radio channels \cite{Wang2017}.} 
%

The clustering problem can be formalized by introducing a set of vectors $\{\bm{\mu}_k, k = 1, 2, \cdots, K\}$, in which $\bm{\mu}_k \in \mathbb{R}^{D \times 1}$ 
represents the center of the $k$th cluster. The motivation is to assign the data set to clusters, such that the distances of each data to its closest cluster center is minimized.
The objective can be rewritten in terms of the total distortion
\begin{equation}
\mathcal{J} = \sum_{p=1}^{\hat{P}} \sum_{k=1}^{K} z_{pk} \left\| \bm{\omega}_{p} - \bm{\mu}_k \right\|^2,
\end{equation}
where $z_{pk} = 1$, if data point $\bm{\omega}_{p}$ is assigned to cluster $k$, otherwise $z_{pk} = 0$.
Each example $\bm{\omega}_{p}$ is assigned or reassigned to its closest cluster center $\mathcal{C}_k$, if
\begin{equation}
\label{eq5ck}
\mathcal{C}_k = \{ n: k = \arg \min_{k} \left\| \bm{\omega}_{p} - \bm{\mu}_k \right\|^2\}.
\end{equation}
The cluster means are updated as
\begin{equation}
\label{eq6um}
\bm{\mu}_k = \frac{1}{|\mathcal{C}_k|}\sum_{p \in  \mathcal{C}_k} \bm{\omega}_{p},
\end{equation}
where $| \cdot |$ is the cardinality of a set, which measures the number of elements of the set. 
The cluster spread is defined as the standard deviation of all the  $\bm{\omega}_{p}$ within the same cluster. 
Recall that all paths within a cluster have the same amplitude, so the mean and spread do not require weighting.
%
Finally, MPC parameter estimates of AOD $(\hat{\theta}_k,\hat{\phi}_k)$, AOA $(\hat{\vartheta}_k,\hat{\varphi}_k)$ and delay $\hat{\tau}_k$ are calculated from spatial frequencies in $\bm{\mu}_k$ as stated in  \fuxi{Sec.~\ref{Sec:TF}}. 

\subsection{Mapping and Localization} 

 \begin{figure}
 \centering
 \includegraphics[width=0.8\columnwidth]{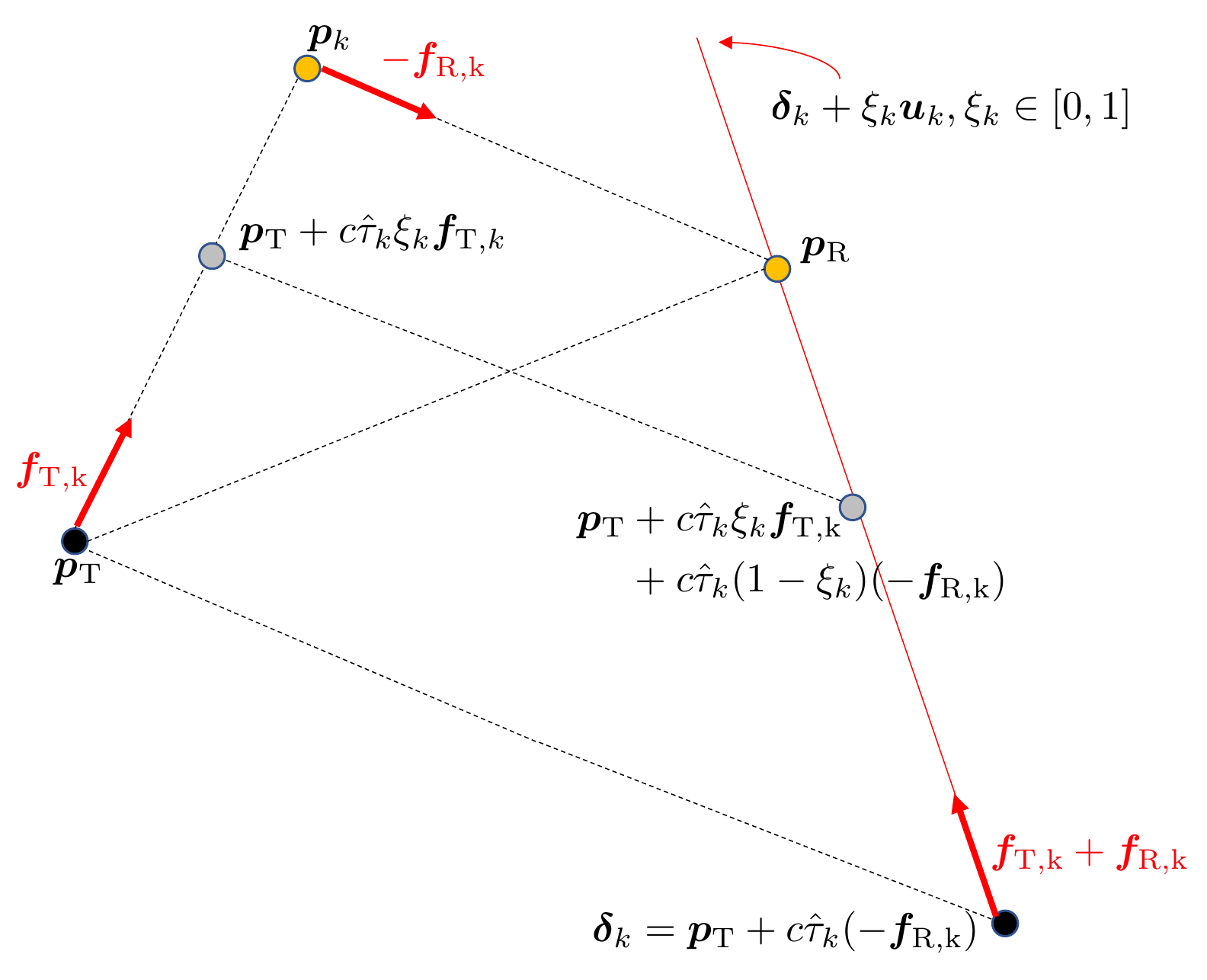}
 \caption{The proposed method for localization and mapping. The locations $\bm{p}_{\mathrm{R}}$ and $\bm{p}_k$ are unknown (in orange). The locations in grey are possible hypothesis of where $\bm{p}_{\mathrm{R}}$ and $\bm{p}_k$ may be, parameterized by $\xi_k \in [0,1]$.}
 \label{fig:PosMethod}
 \end{figure}

We present a general method based on \cite{Wymeersch2018NLOS} that does not rely on knowledge on whether or not the LOS path is present. 
We define
\begin{equation}\label{eqDir}
    \bm{f}_{\mathrm{T},k}=\begin{bmatrix}
       \cos(\hat{\theta}_{k})\sin(\hat{\phi}_{k}) \\ \sin(\hat{\theta}_{k})\sin(\hat{\phi}_{k}) \\ \cos(\hat{\phi}_{k})
     \end{bmatrix},
\end{equation}
which points along the direction of departure of path $k \in \{1,\ldots,\hat{P}\}$; and $\bm{f}_{\rm{R},k}$ is defined equivalently for the direction of arrival.  For each cluster $k$ we can establish a relation to $\bm{p}_{\rm{R}}$ according to
\begin{align}
    \bm{p}_{\rm{R}} = 
 \bm{p}_{\rm{T}} + c\hat{\tau}_k\xi_k \bm{f}_{\mathrm{T},k}  +c\hat{\tau}_k (1-\xi_k)(-\bm{f}_{\mathrm{R},k}),
\end{align}
with unknown $\xi_k \in [0,1]$. Note that for the LOS path (if it is present), the value of $\xi_k$ is arbitrary. In Fig.~\ref{fig:PosMethod}, we show the relation between the different defined vectors and the user location.  Rearranging results in the line equation for each $k$ as
\begin{align}\label{eq-3nlos}
    \bm{p}_{\rm{R}} = \bm{\delta}_k + \xi_k \bm{u}_k,
\end{align}
with $\bm{\delta}_k = \bm{p}_{\rm{T}} -c\hat{\tau}_k \bm{f}_{\mathrm{R},k}$ and $\bm{u}_k = c\hat{\tau}_k( \bm{f}_{\mathrm{T},k}+\bm{f}_{\mathrm{R},k})$. The intersection of these lines determines the estimate of $\bm{p}_{\rm{R}}$. Specifically, we consider the cost function
\begin{align}
    C(\bm{p}_{\rm{R}}) = \sum_{k=1}^{\fuxi{\hat{P}}} \zeta_k \|\bm{p}_{\rm{R}}-(\bm{\delta}_k + \bm{u}_k (\bm{p}_{\rm{R}}-\bm{\delta}_k)^{\mathsf{T}}\bm{u}_k)\|^2, \label{eq:costFunction}
\end{align}
as sum of distance between $\bm{p}_{\rm{R}}$ and each path \eqref{eq-3nlos}, and $\zeta_k \ge 0$ is the weight of the $k$-th path (e.g., dependent on the SNR or the spread of path). 
The least-squares solution becomes
\begin{equation}\label{eq-4rx}
    \hat{\bm{p}}_{\rm{R}} = \left( \sum_{k=1}^{\fuxi{\hat{P}}} \zeta_k(\bm{I}-\bar{\bm{u}}_k\bar{\bm{u}}_k^{\mathsf{T}}) \right)^{-1} \sum_{k=1}^{\fuxi{\hat{P}}} \zeta_k(\bm{I}-\bar{\bm{u}}_k\bar{\bm{u}}_k^{\mathsf{T}})\bm{\delta}_k.
\end{equation}
with $\bar{\bm{u}}_k=\bm{u}_k/\|\bm{u}_k\|$. 

Given $\hat{\bm{p}}_{\rm{R}}$, we can recover the scatter point $\bm{p}_k$ as intersection of the line equations $\bm{p}_{\rm{T}} + \zeta_{\rm{T}}\bm{f}_{\mathrm{T},k}, \zeta_{\rm{T}}\in\mathbb{R}$ and $\bm{p}_{\rm{R}} + \zeta_{\rm{R}}\bm{f}_{\mathrm{R},k}, \zeta_{\mathrm{R}}\in\mathbb{R}$ (see Fig.~\ref{fig:PosMethod}). The least-squares solution follows as
\begin{align}
\hat{\bm{p}}_{k} = (\bm{H}_{\mathrm{T},k}+\bm{H}_{\mathrm{R},k})^{-1}(\bm{H}_{\mathrm{T},k}\bm{p}_{\rm{T}} + \bm{H}_{\mathrm{R},k}\hat{\bm{p}}_{\rm{R}}),
\end{align}
with $\bm{H}_{\mathrm{T},k}=\bm{I}-\bm{f}_{\mathrm{T},k}\bm{f}_{\mathrm{T},k}^{\mathsf{T}}$, $\bm{H}_{\mathrm{R},k}=\bm{I}-\bm{f}_{\mathrm{R},k}\bm{f}_{\mathrm{R},k}^{\mathsf{T}}$ and $\hat{\bm{p}}_{\rm{R}}$ from \eqref{eq-4rx}.

Note that the method does not require separation of specular and diffuse paths. The cost function in \eqref{eq:costFunction} can be applied with all $\hat{P}$ estimated paths, or only a selected 
{subset of paths per cluster. In Section \ref{sec:SLAMLOS}, the performance of different options will be compared.}

\fuxi{In the case multiple users are to be localized simultaneously, the proposed method can be applied independently by each individual user, based on the received downlink signals, as is currently done in LTE. Different levels of cooperation can be envisioned, including map sharing \cite{Kim2020} and exploiting inter-user correlations \cite{Liu2020}.} 

\subsection{Computational complexity}
The most computationally demanding part of channel parameter estimation is {the} CP decomposition.
In general, most CP decomposition algorithms, which factorize $R$-order {tensors},
face high computational cost due to computing
gradients and (approximate) Hessians, line search and rotation.
\fuxi{Table I in \cite{Phan2013} summarizes the complexities of major computations in popular CP decomposition algorithms. For example, the alternating least squares (ALS) algorithm with line search has a complexity of order $\mathcal{O}\left(2^R PJ + RP^3\right)$, where $J = \prod_{r=1}^{R}M_r$ and $P$ denotes the total number of paths. Having $\hat{P}$ multipaths, estimation of $\hat{\boldsymbol{p}}_{\mathrm{R}}$ requires a single $3\times3$ matrix inversion, followed by $\hat{P}+1$ matrix-vector multiplications. In addition, each scatter point estimate demands for a single $3\times3$ matrix inversion plus three matrix-vector multiplications.
Finally,  estimation  of $\hat{\boldsymbol{p}}_{\mathrm{R}}$ requires $\mathcal{O}(\hat{P})$ matrix-vector multiplications.} 
%

\section{Numerical Results}

\subsection{Simulation Setup}

\fuxi{We consider a carrier frequency of 28 GHz, corresponding to $\lambda= 1.07~\text{cm}$, a total bandwidth of 20 MHz with 100 subcarriers, of which 10 equally spaced subcarriers are used for pilots. A cyclic prefix of length 7 is used. 64 pilot OFDM symbols are sent, for a total duration of 3.52 ms. We set the pilots as $\mathbf{S}_i=\mathbf{I}$, $\forall i$. The surface reflection coefficient $\Gamma$ is not specified,  as we only use diffuse paths.}

\begin{figure}
\centering
\input{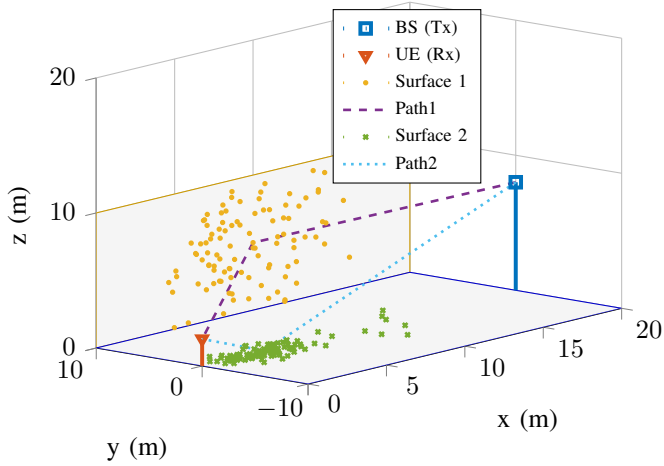}
\caption{Simulation setup for channel estimation and positioning performance evaluation with 2 clusters of different orientation and sizes.}
\label{sim_setup}
\end{figure}

As shown in Fig. \ref{sim_setup},
the transmitter and receiver are located at $\bm{p}_{\rm{T}} = [20, 0, 8]^{\mathsf{T}}$ and 
 $\bm{p}_{\rm{R}} = [0, 0, 2]^{\mathsf{T}}$, respectively, and are surrounded by two surfaces: one building facade and a ground surface. The building facade's center is at $[10,10,5]^{\mathsf{T}}$ with facade length of 20\,m, facade height of 10\,m, and orientation $[0,1,0]^{\mathsf{T}}$ ($x$-$z$ plane). The ground surface is at $[10,0,0]^{\mathsf{T}}$ with orientation $[0,0,1]^{\mathsf{T}}$ (reflected from ground, $x$-$y$ plane), surface dimension is $20\times 20$\,m. 
 Both surfaces are described as rough surfaces \emph{without specular component},
 using $L_k = 100$ scatter points each. 
\fuxi{Furthermore, $K = 2$ is assumed for the following simulations and all the resolved paths are utilized for positioning and mapping,  unless stated otherwise.}
 
The transmitter is equipped with a uniform rectangular array (URA) with ($8 \times 8$) elements and placed along $y$-$z$ plane. 
In both directions, the inter-element spacing is $0.5 \lambda$. 
The origin is the array reference point.
The receiver is also equipped with a URA with ($8 \times 8$) elements and placed along $y$-$z$ plane.

\fuxi{The Matlab package Tensorlab \cite{tensorlab3.0} is utilized for tensor computation, which provides several core algorithms for the computation of the CP decomposition including optimization-based methods such as alternating least squares (ALS), unconstrained nonlinear optimization and nonlinear least squares (NLS). By default, NLS is used for the CP decomposition. It can handle the partially distinct channel parameter scenarios, which was also validated in  \cite{Wen2020}.}


\subsection{Channel Estimation}
We compare the capability of the proposed algorithm to estimate the cluster mean and spread of the multipath parameters using the root-mean-square error (RMSE) for various levels of signal-to-noise ratio (SNR),  defined as
$
\text{SNR} = { \| \mathbfcal{Y} - \mathbfcal{N} \|_{F}^2}/{\| \mathbfcal{N} \|_{F}^2},
$ where $\| \cdot \|_F$ denotes the tensor Frobenius norm \cite{Kolda2009}, and $S$ and $\alpha_R$ are shown in Tables \ref{table1}--\ref{table3}. 
The results are obtained for 100 independent runs.

From Table \ref{table1} (impact of SNR), we observe that the estimation performance improves with SNR. The AOD has a degradation at high SNR, which we attribute to an outlier. The cluster spread estimation also improves with higher SNR. %
In Table \ref{table2} (impact of $S$), we note that when the $S$ parameters increases (more diffuse scattering power), the RMSE performance of the cluster mean and cluster spread both improve. This can be ascribed to more power being available per cluster for a larger value of $S$. 
Finally, Table \ref{table3}  (impact of $\alpha_R$) reveals that when $\alpha_R$ increases (more smooth surface), the RMSE performance of the cluster mean improves, since the paths are more closely clustered around the mean. The  cluster spread RMSE improves somewhat, though the spread itself depends on $\alpha_R$.

 
\begin{table}[!t]
\centering
\caption{RMSE of cluster mean (top) and spread (bottom) - dense components, for various levels of SNR}
\begin{tabular}{c|c c c c} 
 RMSE & \multicolumn{3}{c}{SNR in dB}\\ 
\hline 
$S$ = 0.8, $\alpha_R$ = 10 &  -10 & 0 & 10 \\ 
\hline 
Delay (meter)  &  0.3216  &  0.2025  & 0.1587 \\ 

Azimuth AOD (degree)  & 2.8856  &  2.7250 &  2.1715  \\ 
Elevation AOD (degree)  &  2.6749 & 2.2290 & 3.0189\\ 
  Azimuth AOA (degree)  &  2.0951  & 1.4472   &  1.4027 \\ 
 Elevation AOA (degree)  &   2.4903   &  1.5805   &  1.2407 \\ 
\end{tabular}

\bigskip

\begin{tabular}{c|c c c c} 
 RMSE & \multicolumn{3}{c}{SNR in dB}\\ 
\hline 
$S$ = 0.8, $\alpha_R$ = 10 &  -10 & 0 & 10 \\ 
\hline 
Delay (meter)  & 0.7788   &  0.5835  &  0.4250 \\ 
 Azimuth AOD (degree)  &   9.4723  & 6.3123 &  3.5078 \\
 Elevation AOD (degree)  &   7.4976   &  3.6565   & 2.0295\\
Azimuth AOA (degree)  &    4.0005    &   2.9378    & 1.4150\\
Elevation AOA (degree)  &    4.2699    &   2.1960    &  1.0063\\

\end{tabular}
\label{table1}
\end{table}

\begin{table}[!t]
\centering
\caption{RMSE of cluster mean (top) and spread (bottom) - dense components, for various levels of scatter parameter}
\begin{tabular}{c|c c c c} 
 RMSE & \multicolumn{3}{c}{Scatter Parameter $S$}\\ 
\hline 
  SNR = 10\,dB, $\alpha_R$ = 10  & 0.4 & 0.6 & 0.8 \\ 
\hline 
Delay (meter)   &   0.6774  &  0.2033  &  0.1587\\ 
   Azimuth AOD (degree)  &  2.7009   & 2.4131  &  2.1715  \\
    Elevation AOD (degree)  &  2.8291   & 3.0790  &  3.0189 \\
   Azimuth AOA (degree)  &   2.8190  &  1.5043   & 1.4027 \\
  Elevation AOA (degree)  &      1.9688   &    1.2875   &    1.2407  \\
\end{tabular}

\bigskip

\begin{tabular}{c|c c c c} 
 RMSE & \multicolumn{3}{c}{Scatter Parameter $S$}\\ 
\hline 
 SNR = 10\,dB, $\alpha_R$ = 10 & 0.4 & 0.6 & 0.8 \\ 
\hline 
Delay (meter)     &  0.5078  &  0.4553  &  0.4250  \\ 
Azimuth AOD (degree)      &   4.5017   & 4.9590  &  4.6179  \\ 
Elevation AOD (degree)   &   2.2973  &  2.3543  &  2.1723    \\ 
Azimuth AOA (degree)      &      4.1994   &  1.7163    & 1.7888     \\ 
Elevation AOA (degree)     &    1.2629   & 1.0384   & 1.0303    \\
\end{tabular}
\label{table2}
\end{table}

\begin{table}[!t]
\centering
\caption{RMSE of cluster mean (top) and spread (bottom) - dense components, for various levels of roughness parameter 
}
\begin{tabular}{c|c c c c} 
 RMSE & \multicolumn{3}{c}{Roughness $\alpha_R$}\\ 
\hline 
  SNR = 10\,dB, $S$ = 0.8  & 0 & 10 & 20 \\ 
  \hline 
  Delay (meter)    & 0.4377   & 0.1587  &  0.1224  \\
 
   Azimuth AOD (degree)      &     6.4553  &  2.1715   & 1.8097  \\
  Elevation AOD (degree)     &      3.1283   &  3.0189   & 2.5473   \\
  
  Azimuth AOA (degree)      &      5.2559  &  1.4027  &  0.9615   \\
   Elevation AOA (degree)     &   1.8629  &  1.2407  &  1.1137 \\
\end{tabular}

\bigskip

\begin{tabular}{c|c c c c} 
 RMSE & \multicolumn{3}{c}{Roughness $\alpha_R$}\\ 
\hline 
 SNR = 10\,dB, $S$ = 0.8  & 0 & 10 & 20 \\ 
   \hline 
 Delay (meter)    & 1.0373  &  0.4250  &  0.2694  \\
 
   Azimuth AOD (degree)      &     10.8263  &  4.6179   & 3.5078  \\
  Elevation AOD (degree)     &      2.7932  &  2.1723 &   2.0295  \\
  
  Azimuth AOA (degree)      &      6.0143  &  1.7888  &  1.4150   \\
   Elevation AOA (degree)     &   1.5066  &  1.0303  &  1.0063 \\  

\end{tabular}
\label{table3}
\end{table}


\subsection{Positioning and Mapping in LOS}\label{sec:SLAMLOS}
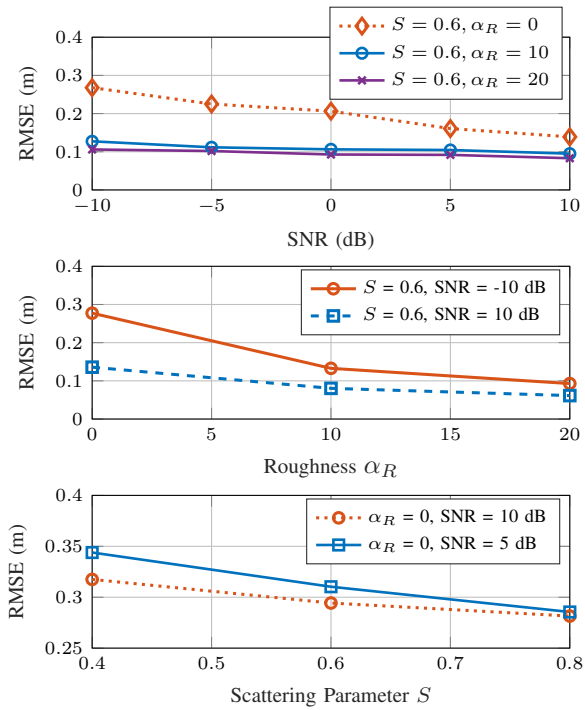
\begin{figure}
\centering
%
%
\definecolor{mycolor1}{rgb}{0.00000,0.44700,0.74100}%
\definecolor{mycolor2}{rgb}{0.85000,0.32500,0.09800}%
\definecolor{mycolor3}{rgb}{0.92900,0.69400,0.12500}%
\definecolor{mycolor4}{rgb}{0.49400,0.18400,0.55600}%
\definecolor{mycolor5}{rgb}{0.46600,0.67400,0.18800}%
\definecolor{mycolor6}{rgb}{0.30100,0.74500,0.93300}%
\definecolor{mycolor7}{rgb}{0.63500,0.07800,0.18400}%
\begin{tikzpicture}

\begin{axis}[%
width=2.5in,
height=0.8in,
at={(0.758in,3.4in)},
scale only axis,
xmin=-10,
xmax=10,
ticklabel style={font=\scriptsize},
xlabel style={font=\color{white!15!black}},
xlabel={\footnotesize{SNR (dB)}},
ymin=0,
ymax=0.4,
ylabel style={font=\color{white!15!black}},
ylabel={\footnotesize{RMSE (m)}},
axis background/.style={fill=white},
xmajorgrids,
ymajorgrids,
legend style={at={(0.985,1.2)}, legend cell align=left, align=left, draw=white!15!black}
]
\addplot [color=mycolor2, line width=1.15pt, mark size=3pt, dotted, mark=diamond, mark options={solid, mycolor2}]
  table[row sep=crcr]{%
-10	0.267853523909452\\
-5	0.225058234410729\\
0	0.206318637220648\\
5	0.160768789729289\\
10	0.139034531486137\\
};
\addlegendentry{\scriptsize{$S = 0.6, \alpha_R = 0$}}

\addplot [color=mycolor1, line width=1.0pt, mark=o, mark options={solid, mycolor1}]
  table[row sep=crcr]{%
-10	0.127363228010355\\
-5	0.111660198864153\\
0	0.106311686737955\\
5	0.104591188032816\\
10	0.0950303924148833\\
};
\addlegendentry{\scriptsize{$S = 0.6,\alpha_R = 10$}}

\addplot [color=mycolor4, line width=1.0pt, mark=x, mark options={solid, mycolor4}]
  table[row sep=crcr]{%
-10	0.106002261071089\\
-5	0.101919005835023\\
0   0.0927830498289708\\
5	0.0918575878272585\\
10	0.0830537711411476\\
};
\addlegendentry{\scriptsize{$S = 0.6,\alpha_R = 20$}}

\end{axis}

\begin{axis}[%
width=2.5in,
height=0.8in,
at={(0.758in,2.2in)},
scale only axis,
xmin=0,
xmax=20,
ticklabel style={font=\scriptsize},
xlabel style={font=\color{white!15!black}},
xlabel={$\footnotesize{\text{Roughness }\alpha_R}$},
ymin=0,
ymax=0.4,
ylabel style={font=\color{white!15!black}},
ylabel={\footnotesize{RMSE (m)}},
axis background/.style={fill=white},
xmajorgrids,
ymajorgrids,
legend style={legend cell align=left, align=left, draw=white!15!black}
]

\addplot [color=mycolor2, line width=1.15pt, mark=o, mark options={solid, mycolor2}]
  table[row sep=crcr]{%
0	 0.2773\\
10	0.1328\\
20	 0.0929\\
};
\addlegendentry{\scriptsize{$S$ = 0.6, SNR = -10 dB}}

\addplot [color=mycolor1, line width=1.15pt, dashed, mark=square, mark options={solid, mycolor1}]
  table[row sep=crcr]{%
0	0.1358 \\
10	0.0804\\
20	0.0614\\
};
\addlegendentry{\scriptsize{$S$ = 0.6, SNR = 10 dB}}

\end{axis}

\begin{axis}[%
width=2.5in,
height=0.8in,
at={(0.758in,1in)},
scale only axis,
xmin=0.4,
xmax=0.8,
ticklabel style={font=\scriptsize},
xlabel style={font=\color{white!15!black}},
xlabel={\footnotesize{Scattering Parameter $S$}},
ymin=0.25,
ymax=0.4,
ylabel style={font=\color{white!15!black}},
ylabel={\footnotesize{RMSE (m)}},
axis background/.style={fill=white},
xmajorgrids,
ymajorgrids,
legend style={legend cell align=left, align=left, draw=white!15!black}
]
\addplot [color=mycolor2, line width=1.15pt, dotted, mark=o, mark options={solid, mycolor2}]
  table[row sep=crcr]{%
0.4	0.3175\\
0.6	0.2943\\
0.8	0.2815\\
};
\addlegendentry{\scriptsize{$\alpha_R$ = 0, SNR = 10 dB}}

\addplot [color=mycolor1, line width=1.0pt, mark=square, mark options={solid, mycolor1}]
  table[row sep=crcr]{%
0.4	0.3439\\
0.6	0.3103\\
0.8	0.2855\\
};
\addlegendentry{\scriptsize{$\alpha_R$ = 0, SNR = 5 dB}}

\end{axis}

\end{tikzpicture}%
\caption{Positioning in LOS \blue{utilizing all  the  resolved  paths}: RMSE versus SNR (top), roughness $\alpha_R$ (middle) and scattering parameter $S$ (bottom).}
\label{rmse_pos_snr}
\end{figure}

The setup is the same as in the Fig.~\ref{sim_setup}, but now it also includes the LOS path. 
%
 %
 %
 Fig.~\ref{rmse_pos_snr} shows the positioning RMSE performance for different SNR, $\alpha_R$ and $S$, with weights $\zeta_k=1$. We observe that thanks to the antenna gains, sub-meter positioning accuracy is achieved when the $\text{SNR} > -10$\,dB. Lower RMSE is achieved with larger scattering parameter $S$. Furthermore, positioning accuracy is sensitive to $\alpha_R$, with more rough surfaces leading to larger RMSE, especially at lower SNRs. 
 
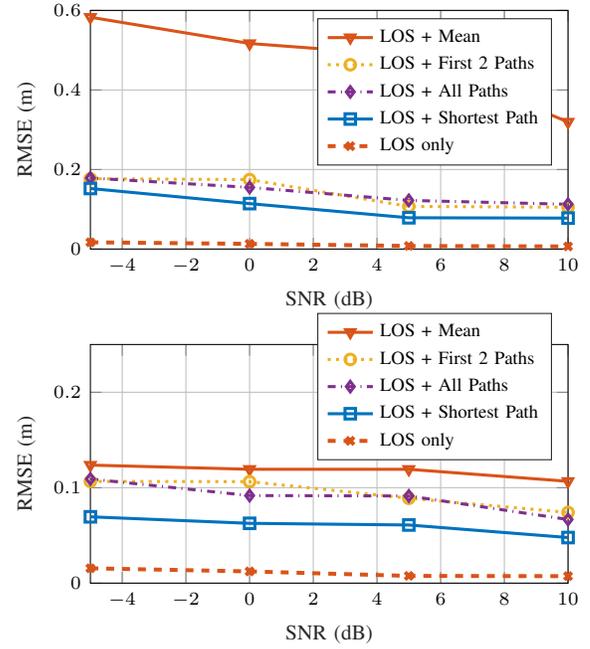
\begin{figure}
\centering
%
%
\definecolor{mycolor1}{rgb}{0.00000,0.44700,0.74100}%
\definecolor{mycolor2}{rgb}{0.85000,0.32500,0.09800}%
\definecolor{mycolor3}{rgb}{0.92900,0.69400,0.12500}%
\definecolor{mycolor4}{rgb}{0.49400,0.18400,0.55600}%
\begin{tikzpicture}

\begin{axis}[%
width=2.5in,
height=1.25in,
at={(0.758in,2in)},
scale only axis,
xmin=-5,
xmax=10,
ticklabel style={font=\scriptsize},
xlabel style={font=\color{white!15!black}},
xlabel={\footnotesize{SNR (dB)}},
ymin=0,
ymax=0.6,
ylabel style={font=\color{white!15!black}},
ylabel={\footnotesize{RMSE (m)}},
axis background/.style={fill=white},
xmajorgrids,
ymajorgrids,
legend style={at={(0.475,0.35)}, anchor=south west, legend cell align=left, align=left, draw=white!15!black,font=\scriptsize}
]

\addplot [color=mycolor2, line width=1.15pt, mark=triangle, mark options={solid, rotate=180, mycolor2}]
  table[row sep=crcr]{%
-5	0.582896024619349\\
0	0.516912869399532\\
5	0.486973942278694\\
10	0.319431304009755\\
};
\addlegendentry{LOS + Mean}

\addplot [color=mycolor3, dotted, line width=1.15pt, mark=o, mark options={solid, mycolor3}]
  table[row sep=crcr]{%
-5	0.176834183338513\\
0	0.174683831300923\\
5	0.107204904993256\\
10	0.105131840860338\\
};
\addlegendentry{LOS + First 2 Paths}

\addplot [color=mycolor4, dashdotted, line width=1.15pt, mark=diamond, mark options={solid, mycolor4}]
  table[row sep=crcr]{%
-5	0.178623521566764\\
0	0.15502580626517\\
5	0.122332155123003\\
10	0.112588225611089\\
};
\addlegendentry{LOS + All Paths}

\addplot [color=mycolor1, line width=1.15pt, mark=square, mark options={solid, mycolor1}]
  table[row sep=crcr]{%
-5	0.152283224952471\\
0	0.11428300732466\\
5	0.07867400659138\\
10	0.0778417000498956\\
};
\addlegendentry{LOS + Shortest Path}

\addplot [color=mycolor2, line width=1.5pt, dashed,mark=x, mark options={solid, rotate=180, mycolor2}]
  table[row sep=crcr]{%
-5	0.016993047956824\\
0	0.0130736773726857\\
5	0.00781556793443202\\
10	0.00656099500861705\\
};
\addlegendentry{LOS only}

\end{axis}

\begin{axis}[%
width=2.5in,
height=1.25in,
at={(0.758in,0.25in)},
scale only axis,
xmin=-5,
xmax=10,
ticklabel style={font=\scriptsize},
xlabel style={font=\color{white!15!black}},
xlabel={\footnotesize{SNR (dB)}},
ymin=0,
ymax=0.25,
ylabel style={font=\color{white!15!black}},
ylabel={\footnotesize{RMSE (m)}},
axis background/.style={fill=white},
xmajorgrids,
ymajorgrids,
legend style={at={(0.475,0.515)}, anchor=south west, legend cell align=left, align=left, draw=white!15!black,font=\scriptsize}
]
\addplot [color=mycolor2, line width=1.15pt, mark=triangle, mark options={solid, rotate=180, mycolor2}]
  table[row sep=crcr]{%
-5	0.123679050821518\\
0	0.119297597356486\\
5	0.119282911323485\\
10	0.10668183043771\\
};
\addlegendentry{LOS + Mean}

\addplot [color=mycolor3, dotted, line width=1.15pt, mark=o, mark options={solid, mycolor3}]
  table[row sep=crcr]{%
-5	0.106575737258727\\
0	0.106336206085552\\
5	0.0889201322772478\\
10	0.0740751629172245\\
};
\addlegendentry{LOS + First 2 Paths}

\addplot [color=mycolor4, dashdotted, line width=1.15pt, mark=diamond, mark options={solid, mycolor4}]
  table[row sep=crcr]{%
-5	0.109131911772778\\
0	0.0917595474504746\\
5	0.0913330370177294\\
10	0.0666972032520153\\
};
\addlegendentry{LOS + All Paths}

\addplot [color=mycolor1, line width=1.15pt, mark=square, mark options={solid, mycolor1}]
  table[row sep=crcr]{%
-5	0.0694918641441355\\
0	0.0626705520095\\
5	0.0609227566946267\\
10	0.0477834210493169\\
};
\addlegendentry{LOS + Shortest Path}

\addplot [color=mycolor2, line width=1.5pt, dashed,mark=x, mark options={solid, rotate=180, mycolor2}]
  table[row sep=crcr]{%
-5	0.0155568598928705\\
0	0.0123140311858798\\
5	0.00760573696626048\\
10	0.00732512719527059\\
};
\addlegendentry{LOS only}
 
\end{axis}
\end{tikzpicture}%
\caption{\fuxi{Positioning in LOS utilizing all  the  resolved  paths: RMSE versus SNR for different algorithms, $S = 0.6$, $\alpha_R = 0$ (top) and $\alpha_R = 10$ (bottom).}}
\label{fig_los_pos_4cases}
\end{figure}

\fuxi{Fig. \ref{fig_los_pos_4cases} shows the positioning performance of utilizing the LOS path and LOS path plus four different combinations of the diffuse paths, which is given by:
\begin{enumerate}
	\item Mean path for each cluster (Mean)
	\item Shortest delay path for each cluster (Shortest Path)
	\item All paths for each cluster (All Paths)
	\item First $2$ paths for each cluster (First 2 Paths).
\end{enumerate}
Compared with the other four algorithms, larger positioning error occurs for the Mean algorithm. That is because just compute the means is not a good approximation of the specular path. 
Lowest RMSE is achieved by only utilizing the LOS path.
That is because when the LOS path is present, the NLOS paths mainly create disturbances, with more diffuse paths leading to larger RMSE.
}

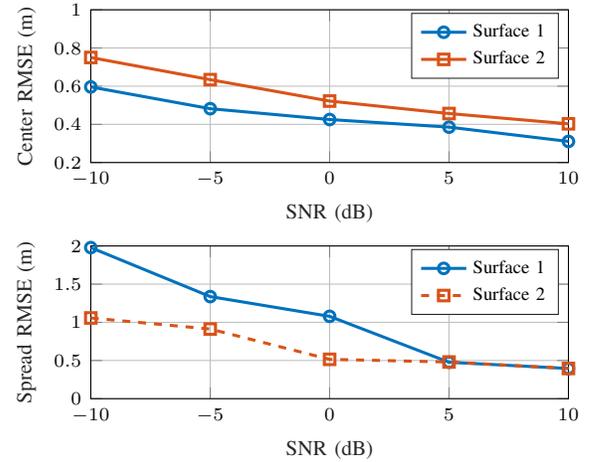
\begin{figure}
\centering
%
%
\definecolor{mycolor1}{rgb}{0.00000,0.44700,0.74100}%
\definecolor{mycolor2}{rgb}{0.85000,0.32500,0.09800}%
\begin{tikzpicture}

\begin{axis}[%
width=2.5in,
height=0.8in,
at={(0.758in,1.8in)},
scale only axis,
xmin=-10,
xmax=10,
ticklabel style={font=\scriptsize},
xlabel style={font=\color{white!15!black}},
xlabel={\footnotesize{SNR (dB)}},
ymin=0.2,
ymax=1,
ylabel style={font=\color{white!15!black}},
ylabel={\footnotesize{Center RMSE (m)}},
axis background/.style={fill=white},
xmajorgrids,
ymajorgrids,
legend style={legend cell align=left, align=left, draw=white!15!black}
]
\addplot [color=mycolor1, line width=1.15pt, mark=o, mark options={solid, mycolor1}]
  table[row sep=crcr]{%
-10	0.595955054284285\\
-5	0.482090890226134\\
0	0.425342175269505\\
5	0.385362987099655\\
10	0.310520016774972\\
};
\addlegendentry{\scriptsize{Surface 1}}

\addplot [color=mycolor2, line width=1.15pt, mark=square, mark options={solid, mycolor2}]
  table[row sep=crcr]{%
-10	0.750204053483216\\
-5	0.63408642741498\\
0	0.522055813192065\\
5	0.456669185183399\\
10	0.402759992508968\\
};
\addlegendentry{\scriptsize{Surface 2}}

\end{axis}

\begin{axis}[%
width=2.5in,
height=0.8in,
at={(0.758in,0.563in)},
scale only axis,
xmin=-10,
xmax=10,
ticklabel style={font=\scriptsize},
xlabel style={font=\color{white!15!black}},
xlabel={\footnotesize{SNR (dB)}},
ymin=0,
ymax=2,
ylabel style={font=\color{white!15!black}},
ylabel={\footnotesize{Spread RMSE (m)}},
axis background/.style={fill=white},
xmajorgrids,
ymajorgrids,
legend style={legend cell align=left, align=left, draw=white!15!black}
]
\addplot [color=mycolor1, line width=1.15pt,solid, mark=o, mark options={solid, mycolor1}]
  table[row sep=crcr]{%
-10	1.97842687345383\\
-5	1.33699449312864\\
0	1.07844358828358\\
5	0.477200352985528\\
10	0.395459854385672\\
};
\addlegendentry{\scriptsize{Surface 1}}

\addplot [color=mycolor2, line width=1.15pt,dashed, mark=square, mark options={solid, mycolor2}]
  table[row sep=crcr]{%
-10	1.05716731966173\\
-5	0.912058079615663\\
0	0.515168880729949\\
5	0.48039237165965\\
10	0.395697497180562\\
};
\addlegendentry{\scriptsize{Surface 2}}

\end{axis}
\end{tikzpicture}%
\caption{Mapping in LOS \blue{utilizing all  the  resolved  paths}: RMSE of estimated center and spread of the reflective surfaces versus SNR, $\alpha_R = 10$ and $S = 0.6$.}
\label{rmse_mapping_snr}
\end{figure}

\begin{figure}
\centering
%
%
\definecolor{mycolor1}{rgb}{0.00000,0.44700,0.74100}%
\definecolor{mycolor2}{rgb}{0.85000,0.32500,0.09800}%
\begin{tikzpicture}

\begin{axis}[%
width=2.5in,
height=0.8in,
at={(0.758in,1.8in)},
scale only axis,
xmin=0,
xmax=20,
ticklabel style={font=\scriptsize},
xlabel style={font=\color{white!15!black}},
xlabel={$\footnotesize{\text{Roughness }\alpha_R}$},
ymin=0,
ymax=1,
ylabel style={font=\color{white!15!black}},
ylabel={\footnotesize{Center RMSE (m)}},
axis background/.style={fill=white},
xmajorgrids,
ymajorgrids,
legend style={at={(0.985,0.925)},legend cell align=left, align=left, draw=white!15!black}
]
\addplot [color=mycolor1, line width=1.15pt, mark=o, mark options={solid, mycolor1}]
  table[row sep=crcr]{%
0	0.5677\\
10	0.2183\\
20	0.1815\\
};
\addlegendentry{\scriptsize{Surface 1}}

\addplot [color=mycolor2, line width=1.15pt, mark=square, mark options={solid, mycolor2}]
  table[row sep=crcr]{%
0	0.8575\\
10	0.3070\\
20	0.2738\\
};
\addlegendentry{\scriptsize{Surface 2}}

\end{axis}

\begin{axis}[%
width=2.5in,
height=0.8in,
at={(0.758in,0.453in)},
scale only axis,
xmin=0,
xmax=20,
ticklabel style={font=\scriptsize},
xlabel style={font=\color{white!15!black}},
xlabel={$\footnotesize{\text{Roughness }\alpha_R}$},
ymin=0,
ymax=2,
ylabel style={font=\color{white!15!black}},
ylabel={\footnotesize{Spread RMSE (m)}},
axis background/.style={fill=white},
xmajorgrids,
ymajorgrids,
legend style={at={(0.675,0.45)}, anchor=south west, legend cell align=left, align=left, draw=white!15!black}
]
\addplot [color=mycolor1, line width=1.15pt, solid,mark=o, mark options={solid, mycolor1}]
  table[row sep=crcr]{%
0	 0.7083\\
10	0.3392\\
20	0.2480\\
};
\addlegendentry{\scriptsize{Surface 1}}

\addplot [color=mycolor2, line width=1.15pt, dashed,mark=square, mark options={solid, mycolor2}]
  table[row sep=crcr]{%
0	 1.8720\\
10	0.3932\\
20	0.3440\\
};
\addlegendentry{\scriptsize{Surface 2}}

\end{axis}
\end{tikzpicture}%
\caption{Mapping in LOS \blue{utilizing all  the  resolved  paths}: RMSE of estimated center and spread of the reflective surfaces versus roughness parameter $S = 0.6$ and SNR = $10$\,dB.}
\label{rmse_mapping_alpha}
\end{figure}
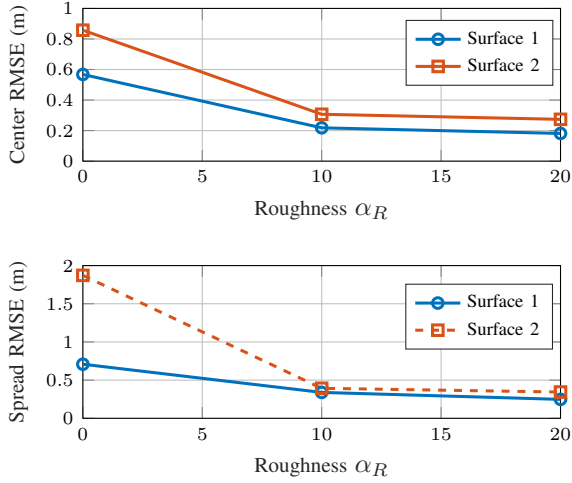

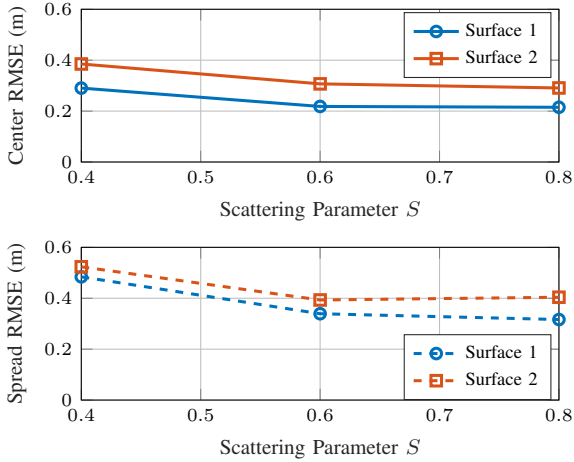
\begin{figure}[!ht]
\centering
%
%
\definecolor{mycolor1}{rgb}{0.00000,0.44700,0.74100}%
\definecolor{mycolor2}{rgb}{0.85000,0.32500,0.09800}%
\begin{tikzpicture}

\begin{axis}[%
width=2.5in,
height=0.8in,
at={(0.758in,1.8in)},
scale only axis,
xmin=0.4,
xmax=0.8,
ticklabel style={font=\scriptsize},
xlabel style={font=\color{white!15!black}},
xlabel={\footnotesize{Scattering Parameter $S$}},
ymin=0,
ymax=0.6,
ylabel style={font=\color{white!15!black}},
ylabel={\footnotesize{Center RMSE (m)}},
axis background/.style={fill=white},
xmajorgrids,
ymajorgrids,
legend style={at={(0.675,0.55)}, anchor=south west, legend cell align=left, align=left, draw=white!15!black}
]
\addplot [color=mycolor1, line width=1.15pt, mark=o, mark options={solid, mycolor1}]
  table[row sep=crcr]{%
0.4	0.2904\\
0.6	0.2183\\
0.8	0.2150 \\
};
\addlegendentry{\scriptsize{Surface 1}}

\addplot [color=mycolor2, line width=1.15pt, mark=square, mark options={solid, mycolor2}]
  table[row sep=crcr]{%
0.4	0.3851\\
0.6	0.3070\\
0.8	0.2909\\
};
\addlegendentry{\scriptsize{Surface 2}}

\end{axis}

\begin{axis}[%
width=2.5in,
height=0.8in,
at={(0.758in,0.553in)},
scale only axis,
xmin=0.4,
xmax=0.8,
ticklabel style={font=\scriptsize},
xlabel style={font=\color{white!15!black}},
xlabel={\footnotesize{Scattering Parameter $S$}},
ymin=0,
ymax=0.6,
ylabel style={font=\color{white!15!black}},
ylabel={\footnotesize{Spread RMSE (m)}},
axis background/.style={fill=white},
xmajorgrids,
ymajorgrids,
legend style={at={(0.675,0.0)}, anchor=south west, legend cell align=left, align=left, draw=white!15!black}
]
\addplot [color=mycolor1, line width=1.15pt, dashed, mark=o, mark options={solid, mycolor1}]
  table[row sep=crcr]{%
0.4	0.4841\\
0.6	0.3392\\
0.8	0.3164\\
};
\addlegendentry{\scriptsize{Surface 1}}

\addplot [color=mycolor2, line width=1.15pt, dashed, mark=square, mark options={solid, mycolor2}]
  table[row sep=crcr]{%
0.4	0.5235\\
0.6	0.3932\\
0.8	0.4040\\
};
\addlegendentry{\scriptsize{Surface 2}}

\end{axis}
\end{tikzpicture}%
\caption{Mapping in LOS \blue{utilizing all  the  resolved  paths}: RMSE of estimated center and spread of the reflective surfaces versus scattering parameter $S$, $\alpha_R = 10$ and SNR = $10$\,dB.}
\label{rmse_mapping_sp}
\end{figure}

The mapping performance is evaluated in terms of the accuracy of the estimated center and spread of the reflective surfaces (clusters). Note that mapping is performed jointly with positioning. The center and spread are defined as the mean and standard deviation of all the estimated $\hat{\bm{p}}_{k}$ within the same cluster.
RMSE of estimated center and spread of the reflective surfaces versus SNR are shown in Fig.~\ref{rmse_mapping_snr}. 
As we expected, high SNR is helpful for center and spread estimation. 
Mapping performance versus scattering parameters  are shown in Fig. \ref{rmse_mapping_alpha}--\ref{rmse_mapping_sp}. Similar to the positioning performance in Fig.~\ref{rmse_pos_snr}, lower RMSE is achieved for larger $\alpha_R$ and larger $S$. 

\begin{figure}
\centering
\includegraphics[width=0.475\textwidth]{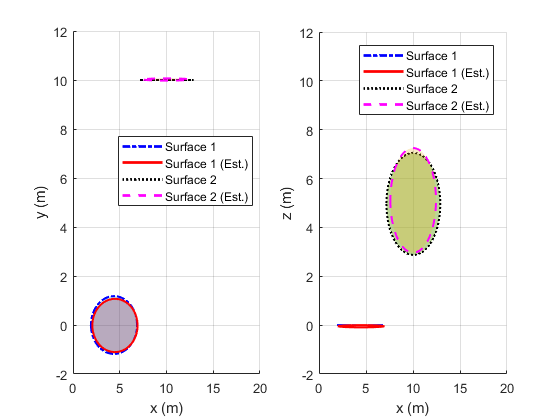}
\caption{Mapping in LOS \blue{utilizing all  the  resolved  paths}: Comparison of the actual and estimated reflective surfaces, projection onto the $x$-$y$ plane (left) and $x$-$z$ plane (right), SNR = 10\,dB, $\alpha_R$ = 10 and $S$ = 0.6.}
\label{figmapping2d}
\end{figure}

The actual and estimated reflective surfaces projected onto the $x$-$y$ plane and $x$-$z$ plane are shown in Fig.~\ref{figmapping2d}. There is a  good match between the  actual and the estimated surface, since both the SNR and $\alpha_R$ value are large. 

\subsection{Positioning and Mapping in NLOS}
 
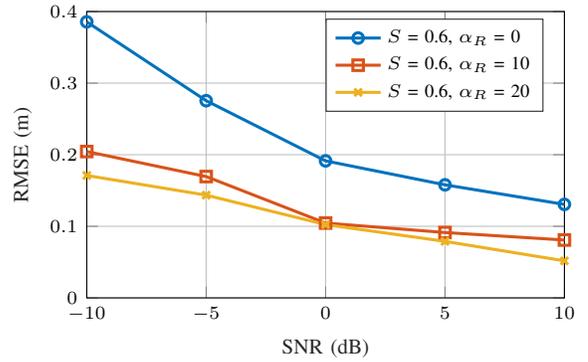
\begin{figure}
\centering
%
%
\definecolor{mycolor1}{rgb}{0.00000,0.44700,0.74100}%
\definecolor{mycolor2}{rgb}{0.85000,0.32500,0.09800}%
\definecolor{mycolor3}{rgb}{0.92900,0.69400,0.12500}%
\begin{tikzpicture}

\begin{axis}[%
width=2.5in,
height=1.5in,
at={(0.758in,1.8in)},
scale only axis,
xmin=-10,
xmax=10,
ticklabel style={font=\scriptsize},
xlabel style={font=\color{white!15!black}},
xlabel={\footnotesize{SNR (dB)}},
ymin=0.0,
ymax=0.4,
ylabel style={font=\color{white!15!black}},
ylabel={\footnotesize{RMSE (m)}},
axis background/.style={fill=white},
xmajorgrids,
ymajorgrids,
legend style={at={(0.5,0.65)}, anchor=south west, legend cell align=left, align=left, draw=white!15!black}
]
\addplot [color=mycolor1, line width=1.15pt, mark=o, mark options={solid, mycolor1}]
  table[row sep=crcr]{%
-10	0.385601666968748\\
-5	0.275459083316383\\
0	0.191415163687722\\
5	0.157943206290588\\
10	0.130609638926327\\
};
\addlegendentry{\scriptsize{$S$ = 0.6, $\alpha_R$ = 0}}

\addplot [color=mycolor2, line width=1.15pt, mark=square, mark options={solid, mycolor2}]
  table[row sep=crcr]{%
-10	0.204468202591217\\
-5	0.169498774476154\\
0	0.104540696898511\\
5	0.0913117971206617\\
10	0.0808509672138093\\
};
\addlegendentry{\scriptsize{$S$ = 0.6, $\alpha_R$ = 10}}

\addplot [color=mycolor3, line width=1.15pt, mark=x, mark options={solid, mycolor3}]
  table[row sep=crcr]{%
-10	0.171016397209103\\
-5	0.143542178502201\\
0	0.102394020947289\\
5	0.0789820607671863\\
10	0.0517562096594721\\
};
\addlegendentry{\scriptsize{$S$ = 0.6, $\alpha_R$ = 20}}

\end{axis}
\end{tikzpicture}
\caption{Positioning in NLOS \blue{utilizing all  the  resolved  paths}: RMSE versus SNR for different values of the scattering parameters. }
\label{rmse_pos_snr_nlos}
\end{figure}

We now move on to the more challenging scenario without LOS. The system setup is the one shown in Fig.~\ref{sim_setup} \blue{ and all the resolved paths are utilized for positioning and mapping}.
Figure~\ref{rmse_pos_snr_nlos} shows the positioning RMSE performance for different SNR and $\alpha_R$.
Similar to LOS scenarios, high SNR is also helpful for positioning in NLOS. Another observation is that lower RMSE is achieved by increasing the roughness parameter $\alpha_R$. 
Overall, performance is somewhat worse than in LOS.

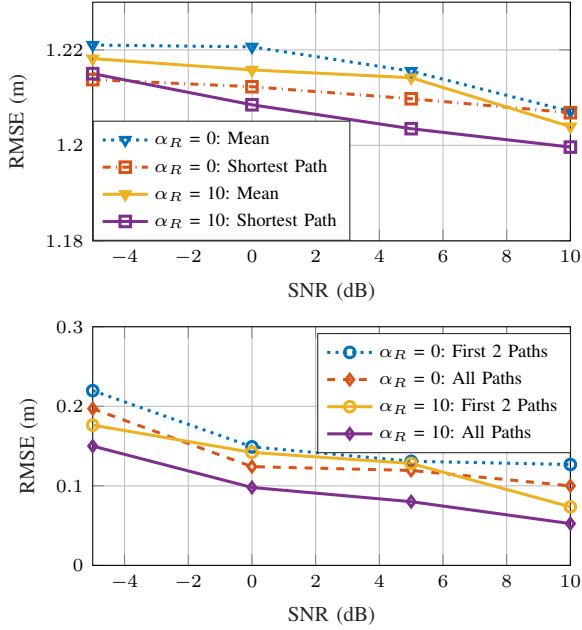
\begin{figure}
\centering
%
%
\definecolor{mycolor1}{rgb}{0.00000,0.44700,0.74100}%
\definecolor{mycolor2}{rgb}{0.85000,0.32500,0.09800}%
\definecolor{mycolor3}{rgb}{0.92900,0.69400,0.12500}%
\definecolor{mycolor4}{rgb}{0.49400,0.18400,0.55600}%
\begin{tikzpicture}

\begin{axis}[%
width=2.5in,
height=1.25in,
at={(0.758in,1.95in)},
scale only axis,
xmin=-5,
xmax=10,
ticklabel style={font=\scriptsize},
xlabel style={font=\color{white!15!black}},
xlabel={\footnotesize{SNR (dB)}},
ymin=1.18,
ymax=1.23,
ylabel style={font=\color{white!15!black}},
ylabel={\footnotesize{RMSE (m)}},
axis background/.style={fill=white},
xmajorgrids,
ymajorgrids,
legend style={at={(0,0)}, anchor=south west, legend cell align=left, align=left, draw=white!15!black,font=\scriptsize}
]

\addplot [color=mycolor1, dotted, line width=1.15pt, mark=triangle, mark options={solid, rotate=180, mycolor1}]
  table[row sep=crcr]{%
-5	1.22101906159068\\
0	1.22064301750757\\
5	1.21554718042125\\
10	1.20711711325309\\
};
\addlegendentry{$\alpha_R$ = 0: Mean}

\addplot [color=mycolor2, dashdotted, line width=1.15pt, mark=square, mark options={solid, mycolor2}]
  table[row sep=crcr]{%
-5	1.2137813373695\\
0	1.21228782200803\\
5	1.20976800596389\\
10	1.20684070661882\\
};
\addlegendentry{$\alpha_R$ = 0: Shortest Path}

\addplot [color=mycolor3,line width=1.15pt, mark=triangle, mark options={solid, rotate=180, mycolor3}]
  table[row sep=crcr]{%
-5	1.21817656517025\\
0	1.21580344047405\\
5	1.21417766721278\\
10	1.20393101972416\\
};
\addlegendentry{$\alpha_R$ = 10: Mean}

\addplot [color=mycolor4,line width=1.15pt, mark=square, mark options={solid, mycolor4}]
  table[row sep=crcr]{%
-5	1.21503063401901\\
0	1.20850561349672\\
5	1.2034964989491\\
10	1.19964355465817\\
};
\addlegendentry{$\alpha_R$ = 10: Shortest Path}

\end{axis}

\begin{axis}[%
width=2.5in,
height=1.25in,
at={(0.758in,0.25in)},
scale only axis,
xmin=-5,
xmax=10,
ticklabel style={font=\scriptsize},
xlabel style={font=\color{white!15!black}},
xlabel={\footnotesize{SNR (dB)}},
ymin=0,
ymax=0.3,
ylabel style={font=\color{white!15!black}},
ylabel={\footnotesize{RMSE (m)}},
axis background/.style={fill=white},
xmajorgrids,
ymajorgrids,
legend style={at={(0.47,0.47)}, anchor=south west, legend cell align=left, align=left, draw=white!15!black,font=\scriptsize}
]

\addplot [color=mycolor1, dotted, line width=1.15pt, mark=o, mark options={solid, mycolor1}]
  table[row sep=crcr]{%
-5	0.219674353493585\\
0	0.149075961238121\\
5	0.13089167339588\\
10	0.126822890588116\\
};
\addlegendentry{$\alpha_R$ = 0: First 2 Paths}

\addplot [color=mycolor2, dashed, line width=1.15pt, mark=diamond, mark options={solid, mycolor2}]
  table[row sep=crcr]{%
-5	0.196968139772471\\
0	0.124120477763634\\
5	0.11934816419425\\
10	0.100031139410523\\
};
\addlegendentry{$\alpha_R$ = 0: All Paths}

\addplot [color=mycolor3, line width=1.15pt, mark=o, mark options={solid, mycolor3}]
  table[row sep=crcr]{%
-5	0.17631880101111\\
0	0.142040471843255\\
5	0.128122002291374\\
10	0.0735937579761125\\
};
\addlegendentry{$\alpha_R$ = 10: First 2 Paths}

\addplot [color=mycolor4, line width=1.15pt, mark=diamond,  mark options={solid, mycolor4}]
  table[row sep=crcr]{%
-5	0.149970229099472\\
0	0.0979586799569517\\
5	0.0800991435880994\\
10	0.0525266326414829\\
};
\addlegendentry{$\alpha_R$ = 10: All Paths}
 
\end{axis}
\end{tikzpicture}%
\caption{\fuxi{Positioning in NLOS: RMSE versus SNR for different algorithms, $S = 0.6$, mean or the shortest path (top) and the first two paths per cluster or all paths (bottom).}}
\label{fig_nlos_pos_4cases}
\end{figure}

\fuxi{Fig. \ref{fig_nlos_pos_4cases} shows the positioning performance in NLOS scenarios. The diffuse paths are helpful to improve the position accuracy. Lower positioning error is achieved by utilizing more diffuse paths and best performance is achieved by using all the estimated diffuse paths. Furthermore, positioning accuracy is sensitive to $\alpha_R$, with more rough surfaces leading to larger RMSE.}

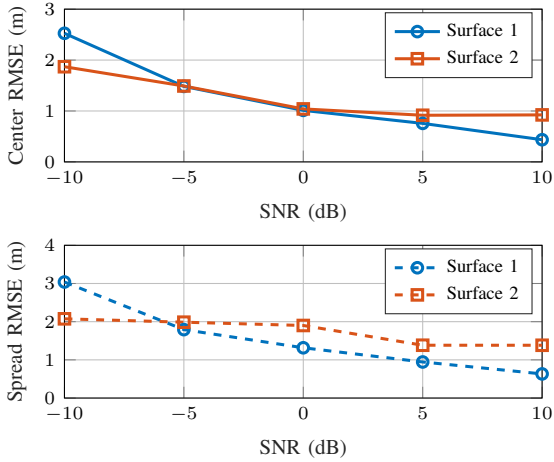
\begin{figure}
\centering
%
%
\definecolor{mycolor1}{rgb}{0.00000,0.44700,0.74100}%
\definecolor{mycolor2}{rgb}{0.85000,0.32500,0.09800}%
\begin{tikzpicture}

\begin{axis}[%
width=2.5in,
height=0.8in,
at={(0.758in,1.8in)},
scale only axis,
xmin=-10,
xmax=10,
ticklabel style={font=\scriptsize},
xlabel style={font=\color{white!15!black}},
xlabel={\footnotesize{SNR (dB)}},
ymin=0,
ymax=3,
ylabel style={font=\color{white!15!black}},
ylabel={\footnotesize{Center RMSE (m)}},
axis background/.style={fill=white},
xmajorgrids,
ymajorgrids,
legend style={legend cell align=left, align=left, draw=white!15!black}
]
\addplot [color=mycolor1, line width=1.15pt, mark=o, mark options={solid, mycolor1}]
  table[row sep=crcr]{%
-10	2.5258\\
-5	1.4830\\
0	1.0120\\
5	0.7563\\
10	0.4355\\
};
\addlegendentry{\scriptsize{Surface 1}}

\addplot [color=mycolor2, line width=1.15pt, mark=square, mark options={solid, mycolor2}]
  table[row sep=crcr]{%
-10	1.8692\\
-5	1.4909\\
0	1.0408\\
5	0.9161\\
10	0.9232\\
};
\addlegendentry{\scriptsize{Surface 2}}

\end{axis}

\begin{axis}[%
width=2.5in,
height=0.8in,
at={(0.758in,0.563in)},
scale only axis,
xmin=-10,
xmax=10,
ticklabel style={font=\scriptsize},
xlabel style={font=\color{white!15!black}},
xlabel={\footnotesize{SNR (dB)}},
ymin=0,
ymax=4,
ylabel style={font=\color{white!15!black}},
ylabel={\footnotesize{Spread RMSE (m)}},
axis background/.style={fill=white},
xmajorgrids,
ymajorgrids,
legend style={legend cell align=left, align=left, draw=white!15!black}
]
\addplot [color=mycolor1, line width=1.15pt,dashed, mark=o, mark options={solid, mycolor1}]
  table[row sep=crcr]{%
-10	3.0390\\
-5	1.7920\\
0	1.3165\\
5	 0.9447\\
10	 0.6319\\
};
\addlegendentry{\scriptsize{Surface 1}}

\addplot [color=mycolor2, line width=1.15pt,dashed, mark=square, mark options={solid, mycolor2}]
  table[row sep=crcr]{%
-10	2.0750\\
-5	1.9867\\
0	1.9004\\
5	1.3833\\
10	1.3825\\
};
\addlegendentry{\scriptsize{Surface 2}}

\end{axis}
\end{tikzpicture}%
\caption{Mapping in NLOS \blue{utilizing all  the  resolved  paths}: RMSE of estimated center and spread of the reflective surfaces versus SNR in NLOS, $\alpha_R = 10$ and $S = 0.6$.}
\label{rmse_mapping_snr_nlos}
\end{figure}

To assess the mapping performance, the RMSE of estimated center and spread of the reflective surfaces versus SNR in NLOS are shown in Fig.~\ref{rmse_mapping_snr_nlos}. Note again that mapping is performed jointly with positioning, so the receiver's position is not known. From Fig.~\ref{rmse_mapping_snr_nlos}, we observe that there is a performance penalty compared to the LOS case, but at sufficiently high SNR, accurate  center  and spread estimates can be obtained. 

\begin{figure}
\centering
\includegraphics[width=0.475\textwidth]{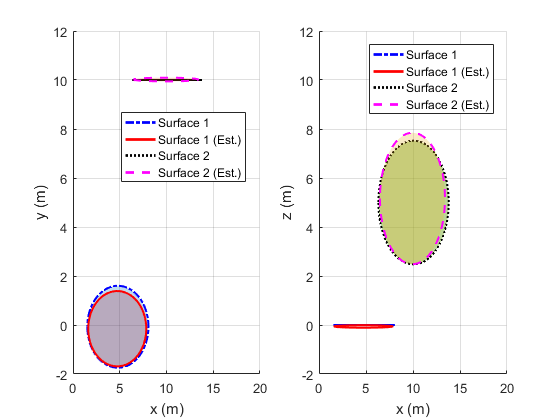}
\caption{Mapping in NLOS \blue{utilizing all  the  resolved  paths}: Comparison of the actual and estimated reflective surfaces in NLOS, projection onto the $x$-$y$ plane (left) and $x$-$z$ plane (right), SNR = $-10$\,dB, $\alpha_R$ = 0 and $S$ = 0.6.}
\label{figmapping2dnlos}
\end{figure}

The actual and estimated reflective surfaces projected onto the $x$-$y$ plane and $x$-$z$ plane in NLOS are shown in Fig.~\ref{figmapping2dnlos}. The mapping error is slightly larger when compared with the LOS scenarios,  because the estimated receiver position is more accurate with LOS. 

\section{Conclusion}
We have studied the problem of channel estimation of mmWave channels with diffuse scattering components, combined with positioning and mapping. 
We proposed a novel  tensor-based  method  for   estimation of the mmWave  channel  parameters  in  a  non-parametric  form. Reflective surfaces with different roughness and scattering parameters are considered. The method   is   able   to   accurately   estimate   the   channel, center and spread of the reflective surfaces,  even   in the  absence  of  a  specular  component.  We also propose a method for localization and mapping based on these channel estimates, and demonstrate that accurate localization of a user and mapping of the environment is possible, even when the LOS path is blocked and surfaces are characterized by only diffuse scattering. 

\appendices
\section{Geometric Relations} \label{sec:appGeometry}
The geometric relations between the location parameters are as follows, with $\bm{p}_{\rm{T}}=[x_{\rm{T}},\,y_{\rm{T}},\,z_{\rm{T}}]^{\mathsf{T}}$,  $\bm{p}_{\rm{R}}=[x_{\rm{R}},\,y_{\rm{R}},\,z_{\rm{R}}]^{\mathsf{T}}$, $\bm{p}_{kl}=[x_{kl},\,y_{kl},\,z_{kl}]^{\mathsf{T}}$:
\begin{itemize}
\item TOA: $\tau_{kl}=\Vert\bm{p}_{kl}-\bm{p}_{\rm{T}}\Vert/c+\Vert\bm{p}_{kl}-\bm{p}_{\rm{R}}\Vert/c$
\item AOA azimuth: $\vartheta_{kl}=\arctan2\left(y_{kl}-y_{\rm{R}},x_{kl}-x_{\rm{R}}\right) + \pi$.
\item AOA elevation: $\varphi_{kl}=\arccos\left((z_{kl}-z_{\rm{R}})/\|\bm{p}_{kl}-\bm{p}_{\rm{R}}\|\right)$.
\item AOD azimuth: $\theta_{kl}=\arctan2\left(y_{kl}-y_{\rm{T}},x_{kl}-x_{\rm{T}}\right)$.
\item AOD elevation: $\phi_{kl}$$=$$\arccos\left((z_{kl}-z_{\rm{T}})/\|\bm{p}_{kl}-\bm{p}_{\rm{T}}\|\right)$.
\end{itemize}

\section{Generation of scatter points and their complex gains} \label{sec:gensp}

Clearly, $p_{\text{DM}}(\bm{p})=0$ for any $\bm{p}$ not lying on a surface. 
To populate the $k$-th rough surface $\mathbb{S}_k$ with scatter points $\{\bm{p}_{kl}\}\subseteq~\mathbb{S}_k, 0<l\leq L_k$, we decompose $p_{\text{DM}}(\bm{p})=\sum_k p_{\text{DM},k}(\bm{p})$ where $p_{\text{DM},k}(\bm{p})$ denotes the JADPS associated to $k$. The JADPS is calculated as function of $\bm{p}_{\rm{T}}$, $\bm{p}_{\rm{R}}$, surface location, as well as $S$ and $\alpha_R$ \cite{kulmer2018impact}. 

To generate the scatter points as well as the complex gains, we consider two methods:
\begin{itemize}
    \item \emph{Rejection sampling:} We generate the scatter points $\bm{p}_{kl}$ such that their density on $\mathbb{S}_k$ is proportional to $p_{\text{DM},k}(\bm{p})$. Then, $p_{\text{DM},k}(\bm{p})$ can be approximated as $\sum_{l=1}^{L_k}|\gamma_{kl}|^2\delta(\bm{p}-\bm{p}_{kl})$ with Dirac delta $\delta$ and $\bm{p}_{kl}$ resulting from {rejection sampling} \cite{Bishop2006}. The corresponding $\gamma_{kl}$ is set equal magnitude $|\gamma_{kl}|=\sqrt{\frac{1}{L_k} P_{k,\text{total}}}$, $P_{k,\text{total}}=\int_{\mathbb{S}_k} p_{\text{DM},k}(\bm{p})\text{d}x\text{d}y\text{d}z$ and random phase, uniform over $[0,2\pi)$.  This procedure allows to describe the JADPS with a rather small $L_k$.
    \item \emph{Uniform sampling:} As an alternative, $\bm{p}_{kl}$ may be distributed uniformly onto $\mathbb{S}_k$. The corresponding $\gamma_{kl}$ are sampled from a zero-mean Normal distribution with variance $p_{\text{DM},k}(\bm{p})$, i.e., $\gamma_{kl}$ has a zero-mean, complex-valued, circularly symmetric gain $\gamma_{kl}$ \cite{besson2000decoupled} with $\mathbb{E}[|\gamma_{kl}|^2]=p_{\text{DM}}(\bm{p}_{kl})\equiv p_{\text{DM}}(\theta_{kl},\phi_{kl},\vartheta_{kl},\varphi_{kl},\tau_{kl})$.  
\end{itemize}
For large surfaces, the large dynamic of the JADPS results in almost zero gain of many $\bm{p}_{kl}$ in the second method, while the first method omits regions in $\mathbb{S}_k$ with small JADPS. Hence, we propose the use of the first method and have used in throughout this current work.

 
\bibliographystyle{IEEEtran}
\bibliography{IEEEabrv,reference}

\end{document}